\documentclass[aps, prd, amsmath, floats, floatfix, twocolumn, superscriptaddress, showpacs]{revtex4-1}
\usepackage{amsmath}    
\usepackage{amssymb}
\usepackage{amsfonts}
\usepackage{amsthm}
\usepackage{braket}
\usepackage{bm}
\usepackage{color}      
\usepackage{dcolumn}
\usepackage[dvipsnames]{xcolor}
\usepackage{epsfig}
\usepackage{graphicx}   
\usepackage{graphics}
\usepackage[latin1]{inputenc}
\usepackage{latexsym}
\usepackage{mathrsfs}
\usepackage{ragged2e}
\usepackage{rotating}
\usepackage{setspace}
\usepackage{siunitx}
\usepackage[caption=false]{subfig}
\DeclareCaptionJustification{justified}{\justifying}
\usepackage{url}
\usepackage{verbatim}	
\usepackage{hyperref}   
\usepackage{float}
\hypersetup{
	colorlinks=true,
	urlcolor=blue,
	citecolor=blue,
	linkcolor=red
}

\usepackage{xspace} 
\usepackage[toc,page]{appendix}
\usepackage{array}

\makeatletter
\newcommand*{\rom}[1]{\expandafter\@slowromancap\romannumeral #1@}

\newcommand{\pix}{{\rm pix}}

\newcommand{\IR}{{\mbox{\tiny IR}}}

\DeclareMathOperator{\sinc}{sinc}
\makeatother
\allowdisplaybreaks

\newcommand{\TAPIR}{Theoretical Astrophysics 350-17, California Institute of Technology, Pasadena, CA 91125, USA}
\newcommand{\WB}{Walter Burke Institute for Theoretical Physics, California Institute of Technology, Pasadena, CA 91125, USA}

\begin{document}
    \title{Quantum Gravity Background in Next-Generation Gravitational Wave Detectors}

    \author{Mathew W.\ Bub}
    \email{mbub@caltech.edu}
    \affiliation{\WB}
    
    \author{Yanbei Chen}
    \email{yanbei@caltech.edu}
    \affiliation{\TAPIR}
    
    \author{Yufeng Du}
    \email{yfdu@caltech.edu}
    \affiliation{\WB}
    
    \author{Dongjun Li}
    \email{dlli@caltech.edu}
    \affiliation{\WB}
    \affiliation{\TAPIR}

    \author{Yiwen Zhang}
    \email{yiwenz@caltech.edu}
    \affiliation{\WB}
    
    \author{Kathryn M.\ Zurek}
    \email{kzurek@caltech.edu}
    \affiliation{\WB}

    \date{\today}
    
    \preprint{CALT-TH-2023-012}
    
    \begin{abstract}
        We study the effects of geontropic vacuum fluctuations in quantum gravity on next-generation terrestrial gravitational wave detectors.  If the VZ effect proposed in Ref.~\cite{Verlinde_Zurek_2019_1}, as modeled in Refs.~\cite{Zurek_2020,Li:2022mvy}, appears in the upcoming GQuEST experiment, we show that it will be a large background for astrophysical gravitational wave searches in observatories like Cosmic Explorer and the Einstein Telescope.
    \end{abstract}

    \maketitle

    \section{Introduction} \label{sec:introduction}

    Bridging the gap between theory and experiment in the study of quantum gravity is at the forefront of research in physics. Although the effects of quantum gravity are ordinarily expected to appear on unobservably-small scales of order the Planck length, $l_p = \sqrt{8 \pi G \hbar / c^3} \sim 10^{-34}~\mathrm{m}$, recent works \cite{Verlinde_Zurek_2019_1, Verlinde_Zurek_2019_2, Zurek_2020, Banks_2021, Gukov:2022oed, Verlinde_Zurek_3,Zhang_2023} have shown that this naive effective field theory (EFT) reasoning may not capture the complete physical picture. Instead, Refs.~\cite{Verlinde_Zurek_2019_1, Verlinde_Zurek_2019_2} showed, using standard holographic techniques, that spacetime fluctuations accumulate from the UV into the IR to produce an effect that scales with the size $L$ of the physical system.   In particular, in flat spacetime, the trajectories of photons in an interferometer of length $L$ enclose a finite spacetime region known as a causal diamond. The geometric fluctuations induced by entropic fluctuations within the causal diamond, or ``geontropic fluctuations,'' manifest as uncertainty in the arm length of the interferometer, as measured by the photon travel time, with a variance that scales as
    \begin{equation} \label{eq:delta_L}
        \langle \Delta L^2 \rangle \sim l_p L.
    \end{equation}
    Additionally, these fluctuations exhibit long-range transverse correlations which enable observation. This result has proven to be theoretically robust, having been confirmed with several distinct theoretical approaches in Refs.~\cite{Zurek_2020, Banks_2021, Gukov:2022oed, Verlinde_Zurek_3,Zhang_2023}, such that the geontropic fluctuations are observed in flat Minkowski, dS, and AdS spacetimes. For a summary of all of these works, see Ref.~\cite{Zurek_2022}.

    More recently, Ref.~\cite{Li:2022mvy}, building upon the work of Ref.~\cite{Zurek_2020}, developed a {\em model} of these geontropic fluctuations in terms of bosonic degrees of freedom coupled to the metric. The model is designed to capture the most prominent features of the theory developed in Refs.~\cite{Verlinde_Zurek_2019_1, Verlinde_Zurek_2019_2, Banks_2021, Gukov:2022oed,Zhang_2023, Verlinde_Zurek_3}, while being local and allowing for the explicit computation of the gauge-invariant interferometer observable.   It features a scalar field $\phi$, the ``pixellon'', a breathing mode corresponding to spacetime fluctuations of the (spherically symmetric) volume of spacetime under observation.  This model allows for the calculation of the power spectral density (PSD) of geontropic fluctuations in spherically-symmetric configurations, in particular for traditional L-shaped interferometers such as LIGO~\cite{Lee_LIGO_2021} and LISA~\cite{LISA_2021}. 
    
    Ref.~\cite{Li:2022mvy} also compared the PSD of the pixellon model to the strain sensitivities of several current and future gravitational wave (GW) detectors, namely LIGO/Virgo~\cite{Lee_LIGO_2021}, Holometer~\cite{Chou_2017}, GEO600~\cite{geo_600}, and LISA~\cite{LISA_2021}. These experiments either produced modest constraints on the pixellon model (in the cases of LIGO and Holometer) or were not sensitive to the model (in the cases of GEO600 and LISA). There are several general reasons for this. For large instruments such as LISA, we expect a reduced signal as the geontropic strain scales parametrically as $h = \frac{\Delta L}{L} \sim \sqrt{\frac{l_p}{L}}$.  On the other hand, existing terrestrial experiments typically have poorer strain sensitivities near the relatively high frequency $\omega_\mathrm{peak} \sim \frac{1}{L}$ at which the pixellon signal achieves its peak. In this paper, we build upon this previous work and survey the landscape of next-generation GW detectors, characterizing their sensitivity to geontropic fluctuations as modeled by the pixellon. We also consider these experiments in the context of the upcoming GQuEST experiment~\cite{mccullerGQuEST}, which explicitly seeks to measure the geontropic signal. Note that in this paper we assume the pixellon is a good {\em physically equivalent} description of the geontropic fluctuations predicted by the VZ effect~\cite{Verlinde_Zurek_2019_1, Verlinde_Zurek_2019_2, Banks_2021, Gukov:2022oed,Zhang_2023,Zurek_2022,Verlinde_Zurek_3}. As discussed above, while it has been shown that the pixellon model reproduces important features of the VZ effect (such as the angular correlations), the physical equivalence in all aspects of the interferometer observable has not been shown, and is the subject of ongoing, first-principles calculations. We plan to update observational signatures as the theoretical modeling captures more aspects of the first-principles calculations.
    
    With this caveat in mind, the paper is organized as follows. In Sec.~\ref{sec:pixellon_model}, we briefly summarize the pixellon model of Refs.~\cite{Zurek_2020, Li:2022mvy}. In Sec.~\ref{sec:HFGW_detectors}, we review a variety of proposed GW detectors following Ref.~\cite{Aggarwal:2020olq}, and discuss their potential sensitivity to the geontropic signal. In Sec.~\ref{sec:pixellon_extension}, we extend the calculation of the pixellon PSD in Ref.~\cite{Li:2022mvy} to more general interferometer-like experiments, particularly for those with geometries other than the traditional L-shape, and for optically-levitated sensors. In Sec.~\ref{sec:interferometers}, we then apply the results to specific experiments and compare the geontropic signal to the expected strain sensitivities of these experiments. Finally, in Sec.~\ref{sec:conclusions}, we collect our results and discuss their implications for the future of GW observation.
    
\begin{figure*}[t]
        \centering
        \subfloat{
            \includegraphics[width=0.48\linewidth]{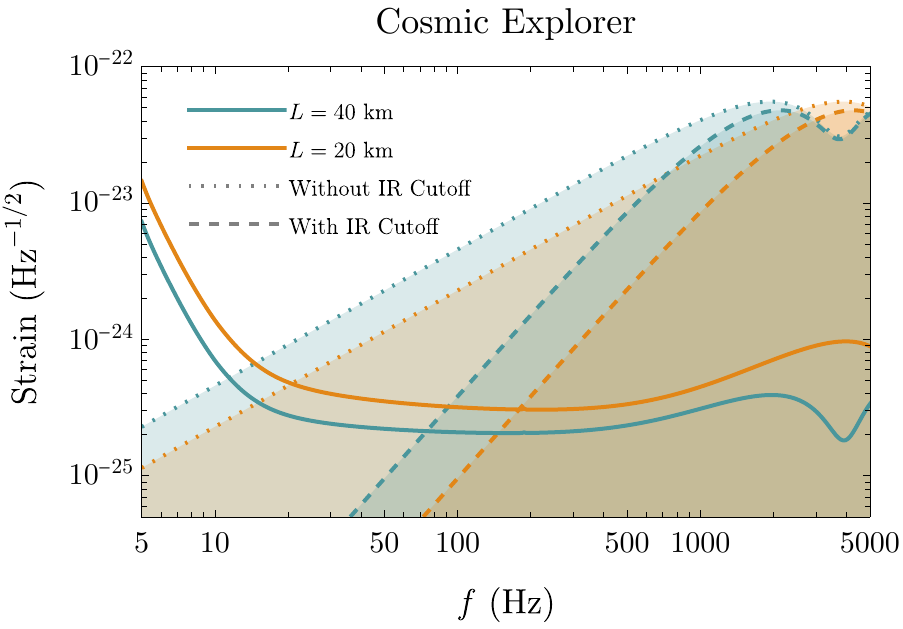}
        }\hfill
        \subfloat{
            \includegraphics[width=0.48\linewidth]{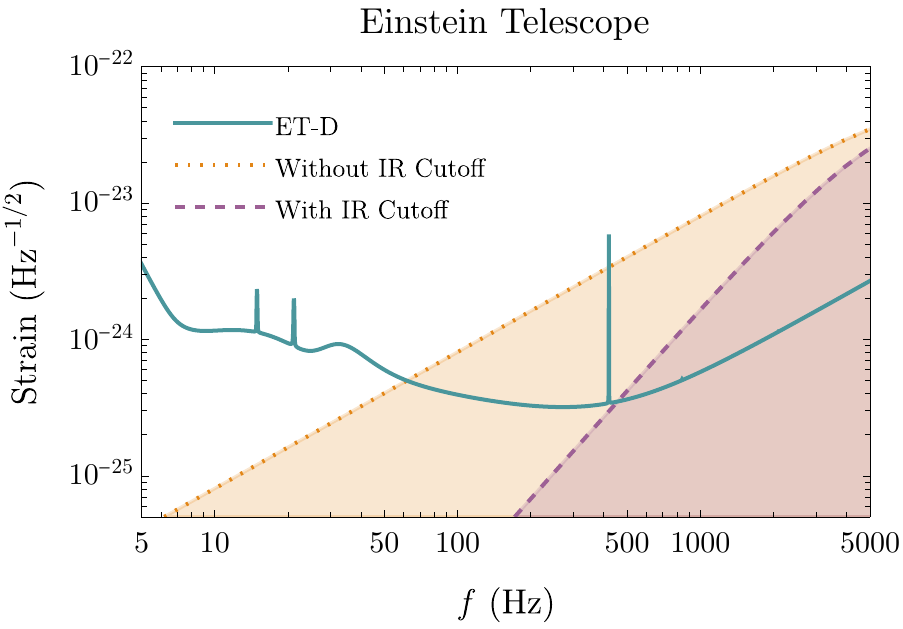}
        }
        \caption{Pixellon strain (dashed and dotted lines) overlaid with the strain sensitivities for CE \cite{Srivastava:2022slt} and ET \cite{Hild:2010id} (solid lines). For CE, we have included both designs with arm lengths $L=20$~km (orange lines) and $L=40$~km (blue lines). The dotted lines give the pixellon strain from Eq.~\eqref{eq:hC} computed without an IR cutoff, and the dashed lines give the same quantity including the IR cutoff from Eq.~\eqref{eq:C_IR_cutoff}. The pixellon strain is computed with the benchmark value $\alpha = 1$.}
        \label{fig:strain_ce_et}
    \end{figure*}

    In anticipation of our main result, in Fig.~\ref{fig:strain_ce_et}, we plot the predicted pixellon signal alongside the strain sensitivities of two prominent next-generation GW detectors: Cosmic Explorer (CE)~\cite{Evans:2021gyd, Srivastava:2022slt} and the Einstein Telescope (ET)~\cite{Hild:2010id}. From these plots, we find a typical geontropic signal exceeds the strain sensitivities of these detectors by two orders of magnitude over a wide range of frequencies. As such, the signal represents a large stochastic background which, if present, would imply a reevaluation of the future of GW astronomy. Moreover, we will show that of the experiments considered in this paper, only CE and ET will have better sensitivity to the geontropic signal than GQuEST, which is a nearer-term apparatus than CE and ET.

    \section{Pixellon Model} \label{sec:pixellon_model}

    In this section, we review the pixellon model proposed in Refs.~\cite{Zurek_2020, Li:2022mvy} to model the geontropic fluctuations of the spherical entangling surface bounding an interferometer, which is also a specialization of the dilaton model studied in Refs.~\cite{Banks_2021, Gukov:2022oed} to causal diamonds in 4-d flat spacetime. Before proceeding, we emphasize that while we expect the pixellon model to reproduce a number of the salient features of the effect proposed in Refs.~\cite{Verlinde_Zurek_2019_1,Verlinde_Zurek_2019_2,Verlinde_Zurek_3,Banks_2021}, the physical equivalence between the model and the complete theory remains to be shown.  Demonstrating this physical equivalence will be crucial for claiming a decisive test of the VZ effect. More specifically, Ref.~\cite{Li:2022mvy} considered a breathing mode of the metric associated with the spacetime volume probed by the interferometer,
    \begin{equation} \label{eq:metric_pix}
		ds^2=-dt^2+(1-\phi)(dr^2+r^2d\Omega^2)\,,
    \end{equation}
    where $\phi$ is a bosonic scalar field,
    \begin{equation} \label{eq:phi_quant}
		\begin{aligned}
		  \phi(x)
		  =& \;l_p\int\frac{d^3\mathbf{p}}{(2\pi)^3}
		  \frac{1}{\sqrt{2\omega(\mathbf{p})}}
		  \left(a_{\mathbf{p}}e^{ip\cdot x}
		  +a_{\mathbf{p}}^{\dagger}e^{-ip\cdot x}\right)\,,
		\end{aligned}
    \end{equation}
    and satisfies the dispersion relation of a sound mode,
     \begin{equation} \label{eq:dispersion}
	    \omega=c_s|\mathbf{p}|\,,\quad
	    c_s=\sqrt{\frac{1}{3}}\,.
    \end{equation}
    The dispersion relation in Eq.~\eqref{eq:dispersion} and the normalization factor $l_p$ in Eq.~\eqref{eq:phi_quant} were derived from plugging the metric in Eq.~\eqref{eq:metric_pix} into the linearized Einstein-Hilbert action \cite{Li:2022mvy}. The creation and annihilation operators $(a_{\mathbf{p}}^{\dagger},a_{\mathbf{p}})$ satisfy the standard commutation relation,
    \begin{equation} \label{eq:commu}
        \left[a_{\mathbf{p}_1},a_{\mathbf{p}_2}^{\dagger}\right]
        =(2\pi)^3\delta^{(3)}(\mathbf{p}_1-\mathbf{p}_2)\,.
    \end{equation}
    
    Instead of being a vacuum state, $\phi$ is thermal with a nontrivial thermal density matrix $\rho_{\pix}$ \cite{Zurek_2020,Li:2022mvy}:
    \begin{align} \label{eq:density_matrix}
		& \rho_{\pix}
		=\frac{1}{\mathcal{Z}}\exp\left[-\beta\int\frac{d^3\mathbf{p}}{(2\pi)^3}
		(\epsilon_{\mathbf{p}}-\mu)
		a_{\mathbf{p}}^{\dagger}a_{\mathbf{p}}\right]\,, \\
		& \mathcal{Z}
		=\prod_{\mathbf{p}}\frac{1}{1-e^{-\beta(\epsilon_{\mathbf{p}}-\mu)}}\,,
    \end{align}
    where $\epsilon_{\mathbf{p}}$ is the energy of the pixellon mode of momentum $\mathbf{p}$, and $\mu$ is the chemical potential counting the background degrees of freedom. In this case, the pixellon modes $\phi$ have an occupation number given by the standard bosonic statistics, {\em i.e.,}
    \begin{align} \label{eq:adagger_a}
        & \operatorname{Tr}\left(\rho_{\pix}
		a_{\mathbf{p}_1}^{\dagger}a_{\mathbf{p}_2}\right)
		=(2\pi)^3\sigma_{\pix}(\mathbf{p}_1)
		\delta^{(3)}(\mathbf{p}_1-\mathbf{p}_2)\,, \nonumber\\
        & \sigma_{\pix}(\mathbf{p})
	    =\frac{1}{e^{\beta(\epsilon_{\mathbf{p}}-\mu)}-1}\,.
    \end{align}
    
    To further simplify the occupation number $\sigma_{\pix}(\mathbf{p})$, Refs.~\cite{Zurek_2020, Li:2022mvy} used that in flat spacetime, the modular Hamiltonian $K$ inside a causal diamond satisfies  \cite{Verlinde_Zurek_2019_1, Banks_2021}
    \begin{equation} \label{eq:flat_DeltaK}
        \langle K\rangle
        \sim\langle\Delta K^2\rangle\sim\frac{A(\Sigma)}{l_p^2}\,,
    \end{equation}
    and similar results in AdS were found in Refs.~\cite{Verlinde_Zurek_2019_2,DeBoer:2018kvc,Nakaguchi:2016zqi}. Since the number of gravitational degrees of freedom $\mathcal{N}$ inside the causal diamond is given by
    \begin{equation}
       \mathcal{N}\equiv\langle K\rangle\,,
    \end{equation}
    the energy fluctuation per degree of freedom is given by \cite{Zurek_2020, Li:2022mvy}
    \begin{equation} \label{eq:energy_momentum}
	    \beta(\epsilon_{\mathbf{p}}-\mu) 
	    \equiv \beta\omega({\bf p})
        \sim\frac{\sqrt{\langle\Delta K^2\rangle}}{\langle K\rangle}
        \sim\frac{l_p}{L}\,.
    \end{equation}
    If one uses Eq.~\eqref{eq:energy_momentum}, identifies $\omega(\mathbf{p})\sim\frac{1}{L}$, and expands $\sigma_{\pix}(\mathbf{p})$ in Eq.~\eqref{eq:adagger_a} to leading order in $\frac{l_p}{L}$, one finds
    \begin{equation} \label{eq:dos}
        \sigma_{\pix}(\mathbf{p})
        =\frac{a}{l_p \omega({\bf p})}\,, 
    \end{equation}
    where $a$ is a dimensionless number, to be fixed by experiment. In Eq.~\eqref{eq:energy_momentum}, $\beta\sim l_p$ corresponds to the local temperature of the near-horizon region probed by the light beams. Comparing the pixellon model here to Refs.~\cite{Verlinde_Zurek_2019_1,Verlinde_Zurek_2019_2,Zurek_2020} and incorporating $\phi$ as a sound mode [{\em i.e.}, Eq.~\eqref{eq:dispersion}], Ref.~\cite{Li:2022mvy} fixed $a=c_s^2/(2\pi)$, which corresponds to $\beta=2\pi l_p/c_s^2$. Defining
    \begin{equation}\label{eq:alpha_a}
        \alpha \equiv \frac{2\pi}{c_s^2}a\,,
    \end{equation}
    we obtain the theory-motivated benchmark for detection $\alpha\sim1$.
  
    In Ref.~\cite{Li:2022mvy}, the pixellon model was used to compute the auto-correlation function of length fluctuations of a single Michelson interferometer with length $L$ and separation angle $\theta$. It was found that the peak of the signal is at $\omega L\sim 1$ with an overall amplitude $\sqrt{\langle \Delta L^2\rangle}\sim\sqrt{l_p L}$. Moreover, the angular correlations from the pixellon model match well with the predictions of Refs.~\cite{Verlinde_Zurek_2019_1, Verlinde_Zurek_3} from shockwave geometry. The peak frequency $\omega_{\mathrm{peak}}\sim\frac{1}{L}$ is consistent with both the identification $\omega(\mathbf{p})\sim\frac{1}{L}$ made by Eq.~\eqref{eq:dos} and the pixellon mode being a breathing mode controlling the size of the spherical entangling surface bounding the interferometer. From this typical frequency and the strain's amplitude, one can directly see that for a general detector probing a causal diamond of size $L$, we need a strain sensitivity $\sqrt{S_h(f)}\lesssim\sqrt{\omega_{\mathrm{peak}}\langle \Delta L^2\rangle}\sim\sqrt{l_p}\sim 10^{-23}~\mathrm{Hz}^{-1/2}$ near the frequency $\omega_{\mathrm{peak}}\sim\frac{1}{L}$, where $S_h(f)$ is the one-sided noise strain defined in Eq.~\eqref{eq:noise_strain_def}. Most current interferometers, especially those aiming for GW detection, do not have such  good strain sensitivity near the free spectral range, which is a higher frequency than is probed by many interferometers. Thus, we would first like to investigate whether other types of high-frequency GW detectors, besides the next-generation interferometers, can potentially detect geontropic signals.

    \section{High-Frequency GW Detectors} 
    \label{sec:HFGW_detectors}
    
    This section follows the review in Ref.~\cite{Aggarwal:2020olq} to investigate a broad class of high-frequency GW detectors with various operating principles. To understand how the detection of geontropic fluctuations fits in this landscape, we first discuss the proposed scientific goals of these high-frequency GW detectors. Most current proposals intend to probe astrophysical objects in unexplored limits, or test quantum gravity near highly curved spacetime. In contrast, the effect considered in Refs.~\cite{Verlinde_Zurek_2019_1, Verlinde_Zurek_2019_2, Zurek_2020, Banks_2021, Gukov:2022oed, Verlinde_Zurek_3, Zurek_2022, Li:2022mvy,Zhang_2023} and this work fills the gap of examining quantum gravity in flat spacetime. Moreover, the necessary sensitivity and frequency range are within the same regime as other science cases, so utilizing these detectors for geontropic signals is natural. In the second half of this section, we examine these detectors' suitabilities for measuring geontropic fluctuations and argue that interferometer-like experiments are the most optimal.

    \subsection{Sources of high-frequency GWs}
    
    Since the successful detection of GWs by the LIGO-Virgo collaboration \cite{LIGOScientific:2016aoc}, there have been continuous efforts to improve the sensitivity of GW detectors at higher frequencies. One direct motivation for this is to study extreme astrophysical objects in limits or environments which cannot be reached by current GW detectors. For example, the merger of sub-solar mass primordial BHs of $10^{-9}$--$10^{-1}\,M_{\odot}$ can emit GWs with frequencies of $10$--$10^9$ kHz \cite{Aggarwal:2020olq}. For neutron stars (NSs), the remnant hot, high-density matter after their merger can generate GWs at either $\sim1$--$4$~kHz \cite{PhysRevD.40.3194} for a BH remnant or $\gtrsim 6$~kHz \cite{Shibata:2006nm,Baiotti:2008ra} for an NS remnant \cite{Ackley:2020atn}. These high-frequency GWs provide opportunities to study different phases of matter predicted by quantum chromodynamics in a high-density finite-temperature environment \cite{Bauswein:2019skm}. At larger scales, high-frequency GW detectors will assist in learning about GWs emitted by the thermal plasma of the early universe \cite{Ghiglieri:2015nfa} ($1$--$100$~GHz), the stochastic GW background generated by primordial BHs \cite{Anantua:2008am} ($10$--$10^{10}$~THz), cosmic strings \cite{Kibble_1976} ($1$--$10^6$~kHz), and other events at cosmological scales \cite{Aggarwal:2020olq}.
    
    One vital application of these high-frequency GW detectors is to explore quantum gravity, the central focus of this work. Standard tests of quantum gravity using GWs focus on examining the properties of quantum BHs against their classical counterparts. For example, GW detections have been used to test the no-hair theorem \cite{Isi:2019aib}, stating that any classical stationary BH (a solution to the Einstein-Maxwell equation) is characterized only by its mass, charge, and angular momentum \cite{Misner:1973prb}. Still, quantum gravity might dress BHs with hair \cite{Giddings:1993de, Berti:2015itd}. The spectrum of GWs can also serve as a test of the horizon's existence \cite{Mark:2017dnq, Du:2018cmp}, where quantum gravity can modify the structure of the near-horizon geometry \cite{Giddings:2016tla}, either drastically via a ``firewall" hiding all quantum effects \cite{Almheiri:2012rt}, or smoothly with the quantum effects extending over some distance around the BH \cite{Giddings:2006sj}. 

    Unlike these standard tests, the series of works in Refs.~\cite{Verlinde_Zurek_2019_1, Verlinde_Zurek_2019_2, Zurek_2020, Banks_2021, Gukov:2022oed, Verlinde_Zurek_3, Zurek_2022, Li:2022mvy, Zhang_2023} instead focus on perturbations of the near-horizon geometry of causal diamonds in flat spacetime due to quantum gravity, which the pixellon models as an effective description. As introduced in Secs.~\ref{sec:introduction} and \ref{sec:pixellon_model} and shown in detail in Sec.~\ref{sec:L-shape}, the length fluctuations induced by the pixellon in an L-shaped interferometer of length $L$ have a size of $\sqrt{\langle\Delta L^2\rangle} \sim \sqrt{l_p L}$ and a peak frequency at $\omega L\sim 1$, corresponding to a PSD with an amplitude of $\sim\sqrt{cl_p}$. For an interferometer, or, more generally, a causal diamond with characteristic size $L\sim10$~m--$10$~km, we need a strain sensitivity of $\sim10^{-23}~\mathrm{Hz}^{-1/2}$ at the peak frequencies of $\frac{1}{L}\sim\,$kHz--MHz, which is within the target sensitivity of many high-frequency GW detectors. Thus, these high-frequency GW detectors planned for various purposes can also be used to test quantum gravity in flat spacetime, which motivates our following investigation.

    \subsection{Detectors for high-frequency GWs}

    \subsubsection{Interferometers}
    The most natural GW detectors to consider are the next-generation interferometers, such as CE \cite{Evans:2021gyd, Srivastava:2022slt}, ET \cite{ET:design}, and NEMO \cite{Ackley:2020atn}, for which the pixellon model was designed to describe the geontropic fluctuations. Although CE and ET are not usually considered high-frequency detectors but instead broadband detectors, they can access frequencies of a few kHz, which are near their free spectral range. For a single interferometer, the causal diamond is naturally defined by the light beams traveling between the mirrors, with its size equal to the interferometer's arm length. Perturbations to the spherical entangling surface bounding the interferometer are then controlled by the pixellon mode. Although the metric in Eq.~\eqref{eq:metric_pix} is not spherically symmetric due to the nontrivial angular dependence of $\phi(x)$, its spatial part is conformal to the metric of a 3-ball, adapting to the spherical symmetry of an interferometer.
    
    The pixellon model and the procedure to compute length fluctuations can be extended to alternative configurations of Michelson interferometers, such as the triangular configuration of ET. In Ref.~\cite{Li:2022mvy}, the PSD and the angular correlations of a single L-shaped interferometer with an arbitrary separation angle were computed. In Sec.~\ref{sec:pixellon_extension}, we further show that the previous results can be extended to multiple interferometers if we consistently correlate pixellons in different causal diamonds. The cross-correlations of different interferometers can then be studied, becoming a smoking gun signature of geontropic signals. Another advantage of studying cross-correlations between detectors is that the cross spectrum of a correlated noise background between different detectors can be detected at a level much lower than their individual independent noise spectra~\cite{allen1999detecting}. 

    One fundamental barrier for an interferometer to reach the high-frequency regime is the quantum shot noise of lasers (or the high uncertainty of the laser's phase quadrature). The most direct solution to this limitation is to increase the laser power $P_{\text{arm}}$, since the PSD of the quantum noise at high frequencies is proportional to $P_{\text{arm}}^{-1/2}$ \cite{Martynov:2019gvu}, which is the approach adopted by NEMO \cite{Ackley:2020atn}. However, increasing laser power is technically challenging, with issues such as the parametric instability of the mirrors' motion due to energy transfer from the light beams \cite{Evans:2015raa} or the thermal deformation of the mirrors \cite{Evans:2021gyd, LIGOScientific:2021kro}. 
    
    Besides increasing laser power, one can also inject squeezed vacuum into the dark port of the interferometer, leading to a reduced phase uncertainty at the cost of sacrificing the sensitivity at low frequencies \cite{Evans:2021gyd}. Nonetheless, Refs.~\cite{Li:2020cwh, Wang:2022ptj} recently proposed that one can connect a quantum parametric amplifier to the interferometer to stabilize the ``white-light cavity" design in Ref.~\cite{Miao:2015pna}, such that the sensitivity at kHz frequencies can be increased without sacrificing the bandwidth. 

   In addition, for detecting a stochastic background like the geontropic signal, which is spatially correlated for two physically overlapping interferometers, a cross-correlation method can be established for each individual detector to dig under shot noise~\cite{Martynov:2017ufx}. This allows us to achieve a better sensitivity than each detector's noise budget for detecting gravitational waves.
    
    Another way to circumvent quantum shot noise is using photon counting instead of the standard homodyne readout \cite{McCuller:2022hum}. Such a readout will be implemented in a proposed $5$~m tabletop interferometer being commissioned by Caltech and Fermilab under the Gravity from the Quantum Entanglement of Space-Time (GQuEST) collaboration \cite{mccullerGQuEST}, which will explicitly target geontropic fluctuations. By employing photon counting and thereby beating the standard quantum limit, GQuEST will be able to place constraints on $\alpha$ substantially more efficiently in terms of integration time than it would with only a homodyne readout. For a detailed examination of the advantages of photon counting, see Ref.~\cite{McCuller:2022hum}. As GQuEST is a tabletop-sized experiment, it will also be capable of probing the angular correlations of the geontropic fluctuations by adjusting its arm angle. Moreover, it is conceived to be a nearer-term instrument than third generation GW detectors such as CE and ET. As such, should the geontropic signal be detected with GQuEST, this information can be incorporated into the design and planning of future GW detectors, whose strain sensitivities to astrophysical signals might be limited by a geontropic background.

    \subsubsection{Optically-levitated sensors}
    
    Besides interferometers, there are other high-frequency GW detectors that operate like an interferometer, such as the optically-levitated sensor described in Refs.~\cite{Arvanitaki:2012cn, Aggarwal:2020umq}. The optically-levitated sensor functions by trapping a dielectric sphere or microdisk in an anti-node of an optical cavity (see Fig.~\ref{fig:levitated_sensor}) \cite{Arvanitaki:2012cn}. One can also build a Michelson interferometer from optically-levitated sensors by inserting the sensors in each arm's cavity (see Fig.~\ref{fig:levitated_sensor_interferometer}) \cite{Aggarwal:2020umq}. As illustrated in Sec.~\ref{sec:levitated_sensor}, one optically-levitated sensor can be effectively treated as two aligned interferometer arms, where the longer arm has the same length $\ell_m$ as the cavity. The shorter arm has length $x_s$, the distance to a chosen anti-node of the trapping field. The optically-levitated sensor measures the differential distance change $\delta\ell_m-\delta x_s$, the correlations of which are similar to an interferometer of length $L=\ell_m-x_s$, but not identical since the two arms have to be treated separately. Moreover, as depicted in Fig.~\ref{fig:levitated_sensor_interferometer}, there are two causal diamonds enclosing the shorter and longer arms, respectively. In Sec.~\ref{sec:pixellon_extension}, we show how to consistently correlate these multiple causal diamonds.
    
    Levitated sensors achieve their gain in sensitivity by making the test masses respond resonantly to gravitational waves whose frequencies match the test masses' natural oscillation frequency in the trapping potential.  In the devices considered by Refs.~\cite{Arvanitaki:2012cn, Aggarwal:2020umq}, sensitivities are mainly constrained by the thermal noise due to heating of the sensor by the scattering light \cite{Aggarwal:2020umq}. The development of techniques to reduce the thermal noise of an optically-trapped object in many other contexts thus allows a better strain sensitivity for the optically-levitated sensor at high frequencies compared to an interferometer \cite{Arvanitaki:2012cn}. It was further found in Ref.~\cite{Aggarwal:2020umq} that by using stacked disks as the sensor, the thermal noise due to photon recoiling can be further reduced. In addition, the high-frequency performance of the levitated sensor is further enhanced by its tunability. Indeed, the experiment achieves its peak strain sensitivity when the trapped object is resonantly excited at the trapping frequency, which is widely tunable via laser intensity \cite{Aggarwal:2020umq}. In Sec.~\ref{sec:levitated_sensor}, we will compare the PSD of length fluctuations measured by the optically-levitated sensor to its predicted strain sensitivity from Ref.~\cite{Aggarwal:2020umq}.

    \subsubsection{Inverse-Gertsenshtein effect and other experiments}
    
    Apart from interferometer-like experiments, there are other high-frequency GW detectors with different working principles. One major class of such experiments uses the inverse-Gertsenshtein effect, which converts gravitons to electromagnetic (EM) waves \cite{gertsenshtein1961a}. For most of these experiments, strong static magnetic fields of several Tesla are used to convert gravitons into photons \cite{Aggarwal:2020olq}. Many of these experiments have been designed to detect ultralight axion dark matter, which can also couple to the EM fields, such as the ones using microwave cavities ({\em e.g.,} ADMX \cite{ADMX:2021nhd}, HAYSTAC \cite{HAYSTAC:2018rwy}, and SQMS \cite{SQMS}) or pickup circuits ({\em e.g}., ABRACADABRA \cite{Kahn:2016aff} and SHAFT \cite{Gramolin:2020ict}) to receive the signal. Refs.~\cite{Ejlli:2019bqj, Berlin:2021txa, Domcke:2022rgu} found that some of these experiments might be sensitive to high-frequency GWs, especially when the geometry of the detector reflects the spin-2 nature of gravitons. For example, Ref.~\cite{Domcke:2022rgu} found that a figure-8 pickup circuit has a much larger sensitivity than a circular loop. For microwave cavities, if the resonant cavity modes have the same spatial profile as the effective current generated by the inverse-Gertsenshtein effect, there is also a boost of the signal \cite{Berlin:2021txa}. 

    The pixellon model considered in Refs.~\cite{Zurek_2020, Li:2022mvy} and this work can be, in principle, used to compute the inverse-Gertsenshtein effect since geontropic fluctuations manifest themselves as metric fluctuations, {\em i.e.}, Eq.~\eqref{eq:metric_pix}. However, in most available calculations, the response to GWs has only been calculated in the transverse-traceless (TT) gauge or the proper detector frame. Moreover, some of these calculations were not careful with gauge invariance. It was recently shown in Ref.~\cite{Berlin:2021txa} that if one incorporates all the physical effects (such as circuits' motion due to coordinate transformation), the observables, such as current density, are gauge invariant. Nonetheless, this proof was done by explicitly computing the observables in these two specific frames without incorporating all possible coordinate transformations.
    
    Such a calculation is usually sufficient for GW detections, but not geontropic fluctuations. First, since geontropic fluctuations have a typical wavelength of the system's size, the long wavelength assumption of the expansion used in the proper detector frame doesn't apply. Second, geontropic fluctuations are not solutions to the vacuum linearized Einstein equations. They cannot be transformed into the TT gauge, despite Eq.~\eqref{eq:metric_pix} being similar to TT gauge, where only light propagation needs to be considered. Thus, one has to be more generous with the frame choices and show that the observables in this type of experiment are invariant under all possible gauge transformations, as Ref.~\cite{Li:2022mvy} demonstrated for length fluctuations in interferometers.

    A more fundamental question is whether the pixellon model is appropriate for describing this type of experiment, especially those using microwave cavities. The pixellon model was designed to effectively describe gravitational perturbations of the spherical entangling surface bounding the interferometer, a spatial slice of the causal diamond defined by the light beams. Within the cavity, there is no freely propagating photon, so the detector doesn't probe the near-horizon geometry of any causal diamond. In this case, the pixellon model might not be a good effective description, and geontropic fluctuations might be negligible since they are driven by near-horizon dynamics \cite{Banks_2021}. Note that the photon counting technique in Ref.~\cite{McCuller:2022hum} also detects the excess photons generated by gravitational perturbations. However, this readout is still embedded in a Michelson interferometer, so there is a well-defined causal diamond.

    Besides the experiments above, other types of high-frequency GW detectors are discussed in Ref.~\cite{Aggarwal:2020olq}, such as the bulk acoustic wave devices \cite{Goryachev:2014yra}, which operate like a resonant mass bar \cite{Weber:1960zz} and measure the vibration of piezoelectric materials due to passing GWs. Similarly, GWs can also deform microwave cavities, which couple different resonant cavity modes and can be detected \cite{Caves:1979kq}. There are also experiments utilizing the coupling between GWs and electron spin, where the collective electron spin excitations or magnons of ferromagnetic crystals due to GWs are measured \cite{Ito:2019wcb, Ito:2022rxn}. Since no causal diamond is being probed in all of these experiments, geontropic signals might be minimal.  For this reason, for the rest of this work, we focus on these interferometer-like experiments and calculate their sensitivity to the pixellon model.  

    \section{Extension of the pixellon model to multiple interferometers} \label{sec:pixellon_extension}

    \begin{figure*}[t]
        \centering
        \subfloat[A single light beam. The beam of length $L$ is sent from $\mathbf{x}_1$ at $t=-L$ to $\mathbf{x}_2$ and then reflected by the end mirror.]{
            \label{fig:single_beam}
            \includegraphics[width=0.3\linewidth]{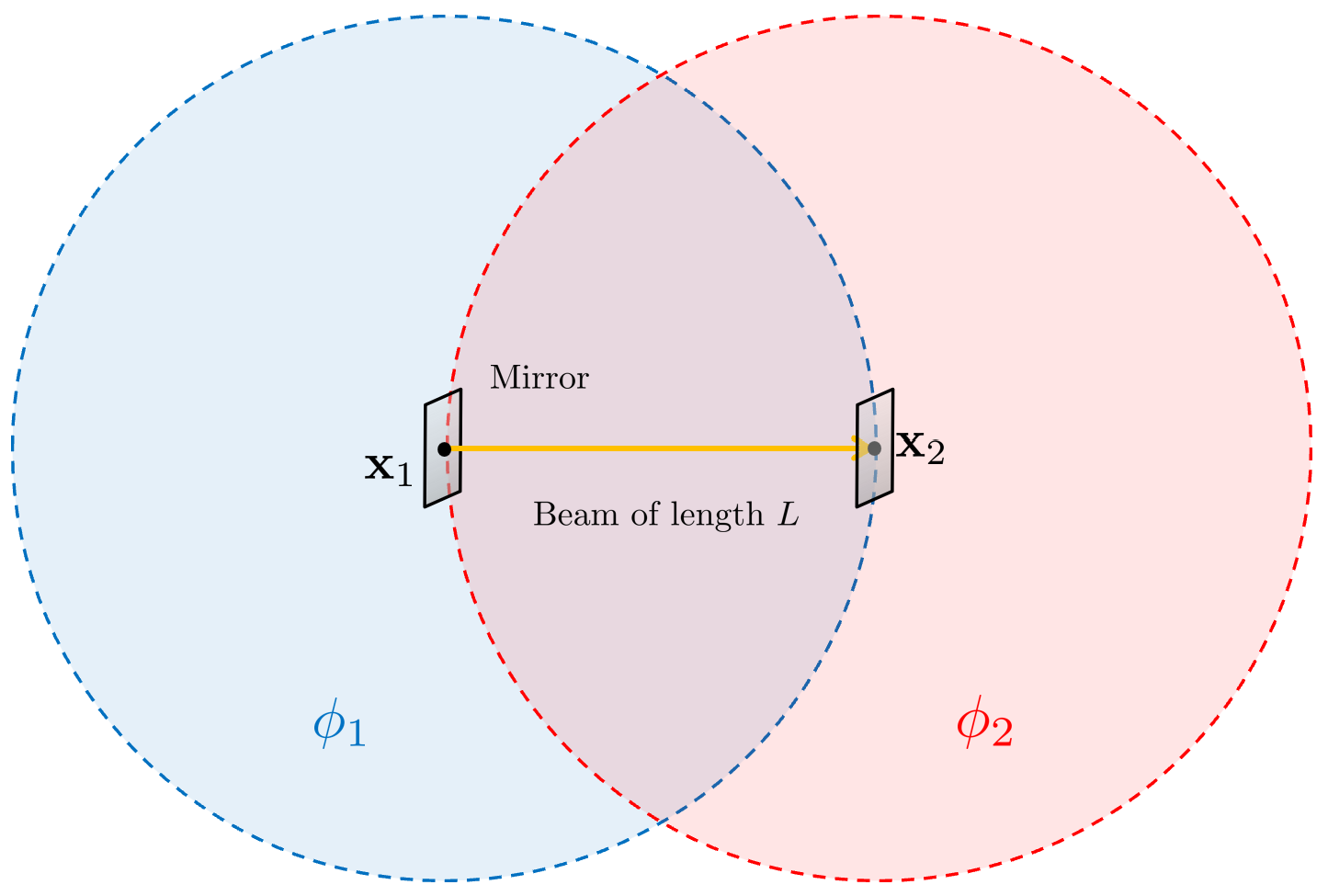}
        }\hfill
        \subfloat[Two light beams. The beam of length $L_1$ is sent from $\mathbf{x}_1$ at $t_1-L_1$ along the direction $\mathbf{n}_1$ and reflected by the end mirror. Similarly, the beam of length $L_2$ is sent from $\mathbf{x}_2$ at $t_2-L_2$ along the direction $\mathbf{n}_2$ and then gets reflected.]{
            \label{fig:two_beams}
            \includegraphics[width=0.3\linewidth]{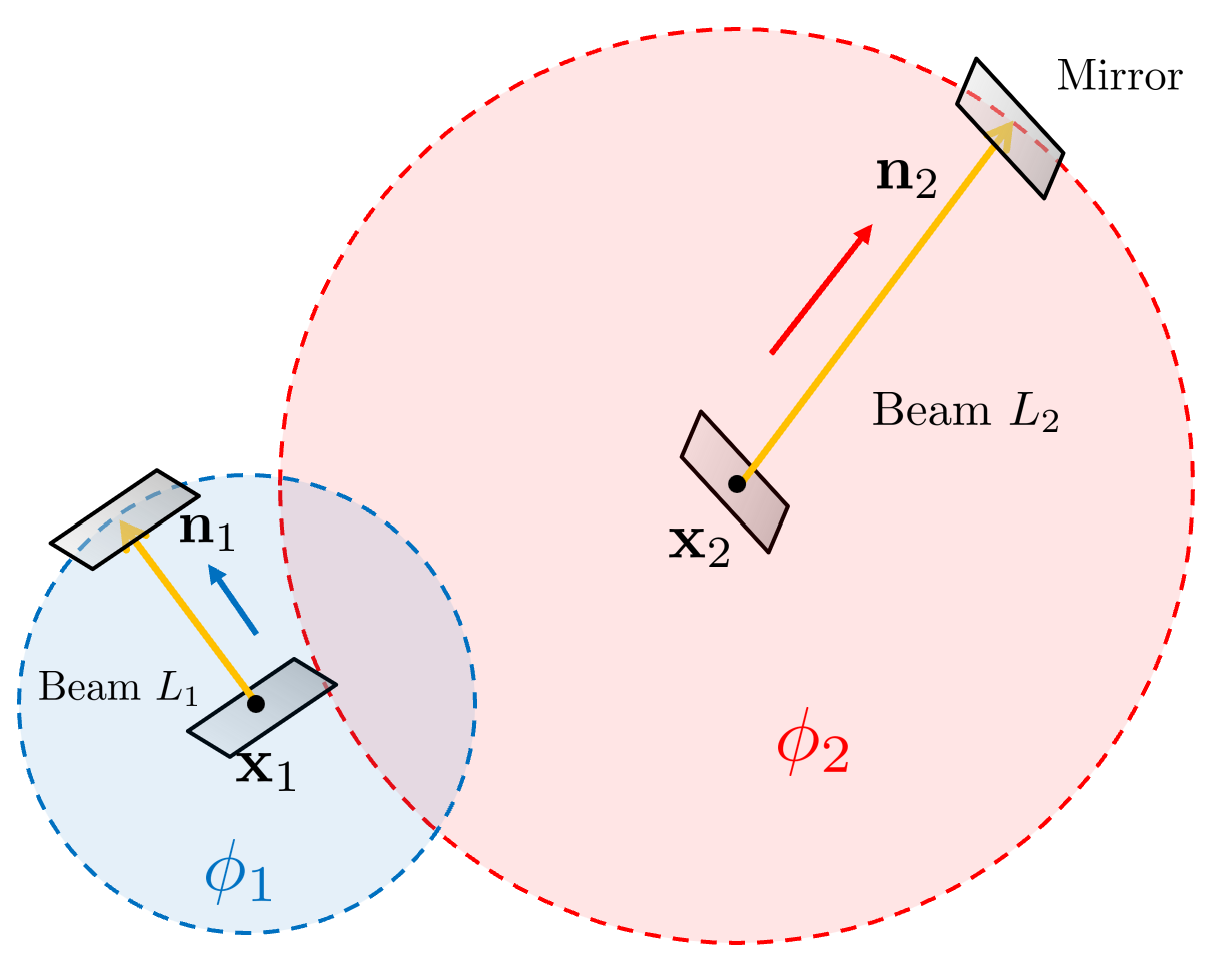}
        }\hfill
        \subfloat[A web of light beams tiling the entire spacetime.]{
            \label{fig:multiple_beams}
            \includegraphics[width=0.3\linewidth]{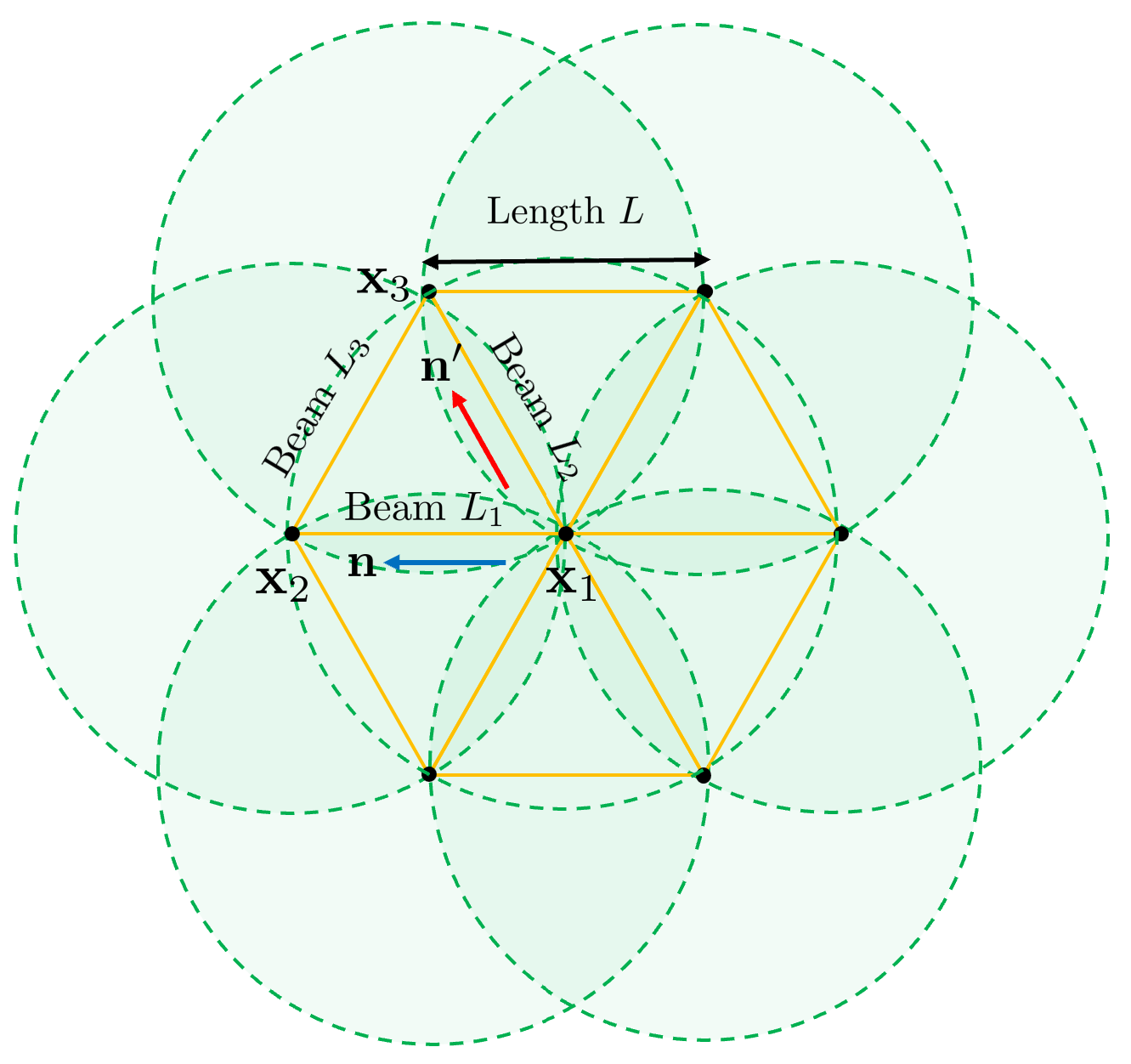}
        }
        \caption{Plots of spherical entangling surfaces or spatial slices of the causal diamonds bounding different configurations of light beams. The shaded circles represent entangling surfaces, each of which is associated with a pixellon model. For all the figures above, we have projected the spherical entangling surface to the plane of the light beams.}
        \label{fig:causal_diamonds}
    \end{figure*}
    
    In this section, we extend the calculation in Ref.~\cite{Li:2022mvy} of the auto-correlation of a single interferometer's length fluctuations to the cross-correlation of two interferometer-like detectors, which may have different arm lengths and origins.

    As shown in Ref.~\cite{Li:2022mvy}, for the metric in Eq.~\eqref{eq:metric_pix}, the only nonzero component in the $t-r$ sector of the metric is $h_{rr}$, so we only need to consider light propagation when computing length fluctuations. For a light beam sent at time $t-L$ from the origin $\mathbf{x}$ along the direction $\mathbf{n}$, the total time delay $T(t,\mathbf{n})$ of a round trip is given by~\footnote{We have corrected a typo in Ref.~\cite{Li:2022mvy}, where the sign before the integral should be minus}
    \begin{align} \label{eq:time_delay_total}
	    & T(t,\mathbf{x}, \mathbf{n})
	    =2L-\frac{1}{2}\int_{0}^{L}dr\;[\phi(x)+\phi(x')]\,, \nonumber \\
	    & x\equiv(t-L+r, \mathbf{x}+r\mathbf{n})\,,\;
	    x'\equiv(t+L-r, \mathbf{x}+r\mathbf{n})\,.
    \end{align}
    Notice that although Eq.~\eqref{eq:time_delay_total} has an explicit dependence on the origin $\mathbf{x}$, the auto-correlation function of $T$ or its fluctuations doesn't depend on $\mathbf{x}$, as shown in Ref.~\cite{Li:2022mvy} and Eq.~\eqref{eq:C_Lshaped_psd}. This indicates that geontropic fluctuations are physical, since they don't depend on the choice of coordinates.
    
    Next, let us consider two light beams sent at times $t_1-L_1$ and $t_2-L_2$ from positions $\mathbf{x}_1$ and $\mathbf{x}_2$ along directions $\mathbf{n}_1$ and $\mathbf{n}_2$, respectively, as depicted in Fig.~\ref{fig:two_beams}. We also assume the lengths of the two beams without any geontropic fluctuations to be $L_1$ and $L_2$, respectively. Then the correlation function of the length fluctuations $\delta T$ of these two beams is
    \begin{align}\label{eq:corr_def}
		& C(\Delta t,\Delta\mathbf{x},\mathbf{n}_{1,2})\equiv
		\left\langle\frac{\delta T(t_1,\mathbf{x}_1,\mathbf{n}_1)\delta T(t_2,\mathbf{x}_2,\mathbf{n}_2)}{4L_1L_2}\right\rangle\,, \nonumber\\
		& \Delta t\equiv t_1-t_2\,,\quad
        \Delta\mathbf{x}\equiv\mathbf{x}_1-\mathbf{x}_2\,,
    \end{align}
    where we have defined $\delta T(t,\mathbf{x},\mathbf{n})=T(t,\mathbf{x},\mathbf{n})-2L$ with $T(t,\mathbf{x},\mathbf{n})$ given in Eq.~\eqref{eq:time_delay_total}. Here, we have assumed that the origins of the light beams enter the cross-correlation function only via their difference $\Delta\mathbf{x}$, so it is independent of the choice of coordinates. We will see this assumption is true in Eq.~\eqref{eq:C_psd}.
    
    Since these two light beams are enclosed by two different causal diamonds as shown in Fig.~\ref{fig:two_beams}, their length fluctuations are separately described by two pixellon models with the metric in Eq.~\eqref{eq:metric_pix} centered at $\mathbf{x}_1$ and $\mathbf{x}_2$, respectively. To distinguish these two pixellon models, we assign $\phi_1(x)$ and $\phi_2(x)$ to the first and the second beams, respectively. Within each pixellon model, the length fluctuations are still described by Eq.~\eqref{eq:time_delay_total}, so
    \begin{equation} \label{eq:dTdT_1}
		\begin{aligned}
		  C(\Delta t,\Delta\mathbf{x},\mathbf{n}_{1,2})
		  =& \;\frac{1}{16L_1L_2}\int_0^{L_1}dr_1\int_{0}^{L_2}dr_2 \\
		  & \;\langle\left(\phi_1(x_1)+\phi_1(x_1')\right)
		  \left(\phi_2(x_2)+\phi_2(x_2')\right)\rangle\,,
		\end{aligned}
    \end{equation}
    which is in a similar form as Eq.~(32) of Ref.~\cite{Li:2022mvy}. For convenience, let us define
    \begin{equation} \label{eq:phi_corr}
        \mathcal{C}(x_1,x_2)
        =\langle(\phi_1(x_1)+\phi_1(x_1'))(\phi_2(x_2)+\phi_2(x_2'))\rangle\,.
    \end{equation}

    To evaluate $\mathcal{C}(x_1,x_2)$, we first need to compute $\langle\phi_1(x_1)\phi_2(x_2)\rangle$, where $x_1$ and $x_2$ are in two different causal diamonds. From Eqs.~\eqref{eq:phi_quant} and \eqref{eq:dispersion}, we notice that both $\phi_1$ and $\phi_2$ satisfy the wave equation, as constrained by the linearized Einstein-Hilbert action \cite{Li:2022mvy}. Thus, $\phi_1$ has translational symmetry, {\em i.e.}, $\phi_1(y)=e^{-ip\cdot(x-y)}\phi_1(x)$ classically, and similarly for $\phi_2$. This implies that although the metric in Eq.~\eqref{eq:metric_pix} effectively describes the length fluctuations of a finite-size interferometer, nothing prevents us from propagating the pixellon field $\phi(x)$ to places outside the interferometer. This is also consistent with the fact that $\phi$ has modes with long wavelengths, as imposed by Eq.~\eqref{eq:dos}. Thus, $\phi_1$ is well-defined in the causal diamond of $\phi_2$, and vice versa.

    To derive a precise relation between $\phi_1$ and $\phi_2$, let us consider a single light beam sent from $\mathbf{x}_1$ at $t=-L$ to $\mathbf{x}_2$, as depicted in Fig.~\ref{fig:single_beam}. To compute the round-trip time delay, one can either use the pixellon model centered at $\mathbf{x}_1$ with the pixellon $\phi_1$, or the one centered at $\mathbf{x}_2$ with the pixellon $\phi_2$. For the former case, we set the origin of the coordinates at $\mathbf{x}_1$ and align the $x$-axis with the outgoing light beam, so the shift of the round-trip time delay $\delta T_1$ is given by Eq.~\eqref{eq:time_delay_total},
    \begin{align} \label{eq:deltaT1}
        & \delta T_1
        =-\frac{1}{2}\int_{0}^{L}dr\;
        \left[\phi_1(x)+\phi_1(x')\right]\,,\nonumber\\
        & x_1=(-L+r,r\hat{\mathbf{x}})\,,\;x_1'=(L-r,r\hat{\mathbf{x}})\,,
    \end{align}
    where the first and second terms correspond to the time delay of the outgoing and ingoing light beams, respectively.
    
    For the latter case, we set the origin at $\mathbf{x}_2$ and align the $x$-axis with the ingoing light beam. Notice the ingoing beam here is the outgoing beam for the pixellon model at $\mathbf{x}_1$, and vice versa. Then, $\delta T_2$ is given by
    \begin{align} \label{eq:deltaT2_1}
        & \delta T_2
        =-\frac{1}{2}\int_{-L}^{0}dr\;
        \left[\phi_2(x)+\phi_2(x')\right]\,,\nonumber\\
        & x_2=(r,r\hat{\mathbf{x}})\,,\;x_2'=(-r,r\hat{\mathbf{x}})\,,
    \end{align}
    where the first and second terms correspond to the time delay of the ingoing and outgoing light beams, respectively.  One can further make a change of variables $\tilde{r}=r+L$ and shift the coordinates, $\mathbf{x}\rightarrow \mathbf{x}+L\hat{\mathbf{x}}$, such that
    \begin{align} \label{eq:deltaT2_2}
        & \delta T_2
        =-\frac{1}{2}\int_{0}^{L}dr\;
        \left[\phi_2(x)+\phi_2(x)\right]\,,\nonumber\\
        & x_2=(-L+r,r\hat{\mathbf{x}})\,,\;x_2'=(L-r,r\hat{\mathbf{x}})\,,
    \end{align}
    where we have replaced the symbol $\tilde{r}$ with $r$ at the end. Since $\delta T_1=\delta T_2$, Eqs.~\eqref{eq:deltaT1} and \eqref{eq:deltaT2_2} indicate that $\phi_1=\phi_2$.

    This relation between $\phi_{1,2}$ does not hold only for these two causal diamonds, but rather the entire spacetime. One can easily see this by tiling the entire spacetime with light beams of the same length $L$ as depicted in Fig.~\ref{fig:multiple_beams}. One can repeat the same argument above for every segment of this web of null rays to relate the pixellon models centered at any two adjacent endpoints. Since all of these null rays are connected, one can easily show a universal $\phi$ across the entire spacetime within the pixellon model. Thus, there is no need to distinguish $\phi$ in different causal diamonds. 

    On the other hand, this does not indicate that we can avoid using separate pixellon models for different light beams. The metric in Eq.~\eqref{eq:metric_pix} is designed to effectively describe the geontropic fluctuations of any causal diamond located at the origin of the local coordinates picked out by the metric. Thus, the light beams not propagating in the radial direction in these local coordinates cannot be described by the associated pixellon model. Furthermore, the argument of gauge invariance of the calculations in Ref.~\cite{Li:2022mvy} does not hold for these non-radial light beams, since the angular directions of the metric were ignored in the proof. Nonetheless, one can always find another causal diamond in which the originally non-radial light beam becomes radial, {\em e.g.}, the causal diamond located at the endpoints of this beam. For example, in Fig.~\ref{fig:multiple_beams}, the beams $L_1$ and $L_2$ can be described by the pixellon model centered at $\mathbf{x}_{1}$, but not the beam $L_3$, although it is in the same causal diamond of the beams $L_{1,2}$. Instead, one should compute the length fluctuations of the beam $L_3$ using the pixellon models at $\mathbf{x}_2$ or $\mathbf{x}_3$. 
    
    One might also worry, in this case, whether the length fluctuations at $\mathbf{x}_1$ have multiple inconsistent descriptions dependent on the causal diamond we choose, particularly with respect to their angular correlations.  For example, since the dominant modes of pixellons are low-$l$ modes \cite{Li:2022mvy}, the pixellon model at $\mathbf{x}_2$ constrains the fluctuations at $\mathbf{x}_1$ to be mostly along $\mathbf{n}$. However, if one uses the pixellon model at $\mathbf{x}_3$, the fluctuations at $\mathbf{x}_1$ are mainly along $\mathbf{n}'$. This is not a contradiction in the pixellon model since light beams in different directions are probing different ``polarizations" of pixellons, which control different local entangling surfaces. If one goes to the causal diamond at $\mathbf{x}_1$, the pixellon model consistently predicts that most fluctuations are along the radial direction, so fluctuations along both $\mathbf{n}$ and $\mathbf{n}'$ can potentially be excited. When the light beam is sent along one of these directions, the spherical symmetry is broken by exciting fluctuations mainly in this specific direction.
     
    In this case, to compute the correlation of any two beams as depicted in Fig.~\ref{fig:two_beams}, we use the metric in Eq.~\eqref{eq:metric_pix} centered at $\mathbf{x}_1$ for beam $L_1$ and the one at $\mathbf{x}_2$ for beam $L_2$, but do not distinguish $\phi$ in these two metrics. Thus, Eq.~\eqref{eq:phi_corr} becomes
    \begin{equation} \label{eq:phi_corr2}
        \mathcal{C}(x_1,x_2)
        =\langle(\phi(x_1)+\phi(x_1'))(\phi(x_2)+\phi(x_2'))\rangle\,.
    \end{equation}
    Using Eq.~\eqref{eq:phi_quant}, we get
    \begin{align} \label{eq:phi_corr_ab}
        \mathcal{C}(x_1,x_2)
        =& \;4l_p^2 \int\frac{d^3 \mathbf{p}_1}{(2\pi)^3} 
        \int\frac{d^3\mathbf{p}_2}{(2\pi)^3}
        \frac{1}{\sqrt{4\omega_1(\mathbf{p}_1)\omega_2(\mathbf{p}_2)}} \nonumber\\ 
        & \;\cos[\omega_1(L_1-r_1)]\cos[\omega_2(L_2-r_2)]
        \big[\langle a_{\bm{p}_1}a_{\bm{p}_2}^\dagger\rangle\nonumber\\
        & \;e^{-i[\omega_1 t_1-\omega_2 t_2
        -\mathbf{p}_1\cdot(\mathbf{x}_1+r_1\mathbf{n}_1)
        +\mathbf{p}_2\cdot(\mathbf{x}_2+r_2\mathbf{n}_2)]}
        +c.c.\big]\,, \nonumber\\
        =& \;4l_p^2\int\frac{d^3 \mathbf{p}}{(2\pi)^3}
        \frac{\sigma_{\pix}(\mathbf{p})}{2\omega(\mathbf{p})} \cos[\omega(L_1-r_1)] \nonumber\\ 
        & \;\cos[\omega(L_2-r_2)]
        \left[e^{-i\omega\Delta t+i\mathbf{p}\cdot
        \delta \mathbf{x}}+c.c.\right]\,,
    \end{align}
    where we have defined
    \begin{equation}
        \delta\mathbf{x}\equiv
        \Delta\mathbf{x}+r_1\mathbf{n}_1-r_2\mathbf{n}_2\,.
    \end{equation}
    Plugging the occupation number in Eq.~\eqref{eq:dos}, the correlation function of the length fluctuations is given by
    \begin{equation} \label{eq:dTdT_3}
		\begin{aligned}
            & C(\Delta t,\Delta\mathbf{x},\mathbf{n}_{1,2}) \\
		  =& \;\frac{al_p}{8L_1L_2}
		  \int_{0}^{L_1}dr_1\int_{0}^{L_2}dr_2\;
		  \int\frac{d^3\mathbf{p}}{(2\pi)^3}
		  \frac{1}{\omega^2(\mathbf{p})} \\
		  & \;\cos{\left[\omega(L_1-r_1)\right]}
		  \cos{\left[\omega(L_2-r_2)\right]}
		  e^{-i\omega\Delta t+i\mathbf{p}\cdot\delta\mathbf{x}}\,,
		\end{aligned}
    \end{equation}
    where we have dropped the $c.c.$ term and hereafter assume for simplicity that the complex conjugate is included implicitly.
    
    Eq.~\eqref{eq:dTdT_3} is very similar to Eq.~(41) of Ref.~\cite{Li:2022mvy}, except that $\delta \mathbf{x}$ also contains the difference between the origins of the two light beams. Evaluating the angular part of the momentum integral, we have
    \begin{equation} \label{eq:corr}
	\begin{aligned}
		  & C(\Delta t,\Delta\mathbf{x},\mathbf{n}_{1,2}) \\
		  =& \;\frac{al_p}{16\pi^2c_s^3L_1L_2}
            \int_{0}^{L_1}dr_1\int_{0}^{L_2}dr_2\int_{0}^{\infty} d\omega\; \\
		  & \;\cos{\left[\omega(L_1-r_1)\right]}
		  \cos{\left[\omega(L_2-r_2)\right]} \\
		  & \;\sinc\left[\omega\mathcal{D}             (r_{1,2},\Delta\mathbf{x},\mathbf{n}_{1,2})/c_s\right]
            e^{-i\omega\Delta t} \,,
	\end{aligned}
    \end{equation}
    with
    \begin{equation}
        \mathcal{D}(r_{1,2},\Delta\mathbf{x},\mathbf{n}_{1,2})=
        |\delta\mathbf{x}|\,.
    \end{equation}
    The PSD $\tilde{C}(\omega,\Delta\mathbf{x},\mathbf{n}_{1,2})$ is then given by
    \begin{equation} \label{eq:C_psd}
	\begin{aligned}
		  & \tilde{C}(\omega,\Delta\mathbf{x},\mathbf{n}_{1,2}) \\
		  =& \;\frac{al_p}{8\pi c_s^3N}
		  \int_{0}^{L_1}dr_1\int_{0}^{L_2}dr_2\;
		  \cos{\left[\omega(L_1-r_1)\right]}\\
		  & \cos{\left[\omega(L_2-r_2)\right]}
            \sinc\left[\omega\mathcal{D}  (r_{1,2},\Delta\mathbf{x},\mathbf{n}_{1,2})/c_s\right]\,,
	\end{aligned}
    \end{equation}
    where we have absorbed the normalization $L_1L_2$ into $N$. We make this redefinition for convenience since in certain experiments discussed later, PSDs similar to Eq.~\eqref{eq:C_psd} appear but with $N\neq L_1L_2$. If we also insert an IR cutoff $\omega^2(\mathbf{p})\rightarrow\omega^2(\mathbf{p})+\omega_{\IR}^2$ in Eq.~\eqref{eq:dTdT_3} similar to Ref.~\cite{Verlinde_Zurek_2019_1}, it was found in Ref.~\cite{Li:2022mvy} that
    \begin{equation} \label{eq:C_IR_cutoff}
        \tilde{C}(\omega,\Delta\mathbf{x},\mathbf{n}_{1,2})\rightarrow
        \frac{\omega^2}{\omega^2+\omega_{\IR}^2}\tilde{C}(\omega,\Delta\mathbf{x},\mathbf{n}_{1,2})\,.
    \end{equation}
    In the case that the two arms have the same length $L$, Ref.~\cite{Li:2022mvy} fixed $\omega_{\IR}=\frac{1}{L}$, which gave a better agreement with the angular correlations predicted in Refs.~\cite{Verlinde_Zurek_2019_1, Verlinde_Zurek_3}.

    One direct application of the results above is to compute the cross-correlation of length fluctuations across two different interferometers. Let the origins of two interferometers be at $\mathbf{x}_{I,II}$, respectively. For the interferometer at $\mathbf{x}_I$, let its two arms be along the directions $\mathbf{n}_{1,2}$ with length $L_{I}$. Similarly, let the two arms of the interferometer at $\mathbf{x}_{II}$ be along the directions $\mathbf{n}_{3,4}$ with length $L_{II}$. Define $\mathcal{T}(\mathbf{x},t)$ to be the difference of length fluctuations of two arms within a single interferometer at position $\mathbf{x}$, the light beams of which are sent at time $t$. Then the cross-correlation of the time difference across two arms is
    \begin{align}\label{eq:corr_T_def}
		& C_{\mathcal{T}}(\Delta t,\Delta\mathbf{x},\mathbf{n}_{I,II})
		\equiv\left\langle\frac{\mathcal{T}_{I}(\mathbf{x}_I,t_1)\mathcal{T}_{II}(\mathbf{x}_2,t_2)}{4L_{I}L_{II}}\right\rangle\,, \nonumber\\
		& \mathcal{T}_{I}(\mathbf{x}_{I},t_1)
		= \delta T(t_I,\mathbf{x}_{I},\mathbf{n}_2)
        -\delta T(t_I,\mathbf{x}_{I},\mathbf{n}_1)\,,
        \nonumber\\
        & \mathcal{T}_{II}(\mathbf{x}_{II},t_2)
		= \delta T(t_{II},\mathbf{x}_{II},\mathbf{n}_4)
        -\delta T(t_{II},\mathbf{x}_{II},\mathbf{n}_3)\,,
    \end{align}
    where $\mathbf{n}_{I}=(\mathbf{n}_1,\mathbf{n}_2)$, $\mathbf{n}_{II}=(\mathbf{n}_3,\mathbf{n}_4)$, and $\Delta\mathbf{x}=\mathbf{x}_{I}-\mathbf{x}_{II}$ such that
    \begin{equation} \label{eq:tautau}
		\begin{aligned}
		  & \tilde{C}_{\mathcal{T}}(\omega,\Delta\mathbf{x},   \mathbf{n}_{I,II}) \\
		  =& \;\tilde{C}(\omega,\Delta\mathbf{x},\mathbf{n}_{1,3})
            +\tilde{C}(\omega,\Delta\mathbf{x},\mathbf{n}_{2,4}) \\
            & \;-\tilde{C}(\omega,\Delta\mathbf{x},\mathbf{n}_{1,4})
            -\tilde{C}(\omega,\Delta\mathbf{x},\mathbf{n}_{2,3})\,.
		\end{aligned}
    \end{equation}
    The equation above generally contains complicated geometric factors, and the integral within Eq.~\eqref{eq:corr} cannot be easily evaluated for a generic geometry. Thus, we consider several specific configurations in the next section.

    \section{Interferometer-like experiments} \label{sec:interferometers}
    
    In this section, we apply the results of Sec.~\ref{sec:pixellon_extension} to several types of interferometer-like experiments: a single L-shaped interferometer ({\em e.g.}, LIGO \cite{Lee_LIGO_2021}, CE \cite{Evans:2021gyd, Srivastava:2022slt}, NEMO \cite{Ackley:2020atn}), the equilateral triangle configuration of multiple interferometers ({\em e.g.}, LISA \cite{LISA_2021}, ET \cite{Hild:2010id}), and optically-levitated sensors \cite{Arvanitaki:2012cn, Aggarwal:2020umq}. 
    \subsection{Single L-shaped interferometer}
    \label{sec:L-shape}

    \begin{figure*}[t]
        \centering
   {
            \includegraphics[width=0.48\linewidth]{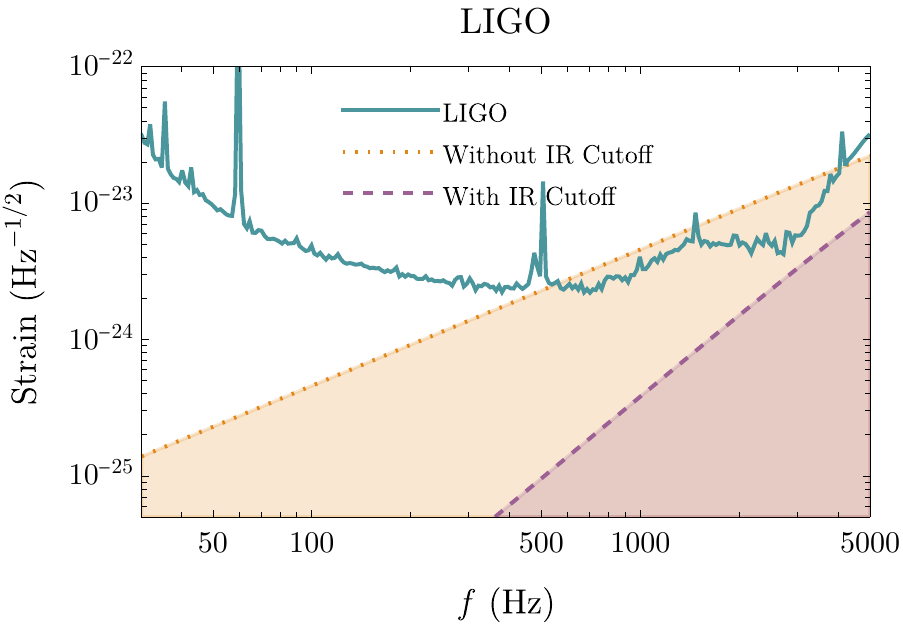}
        }\hfill
        \subfloat{
            \includegraphics[width=0.48\linewidth]{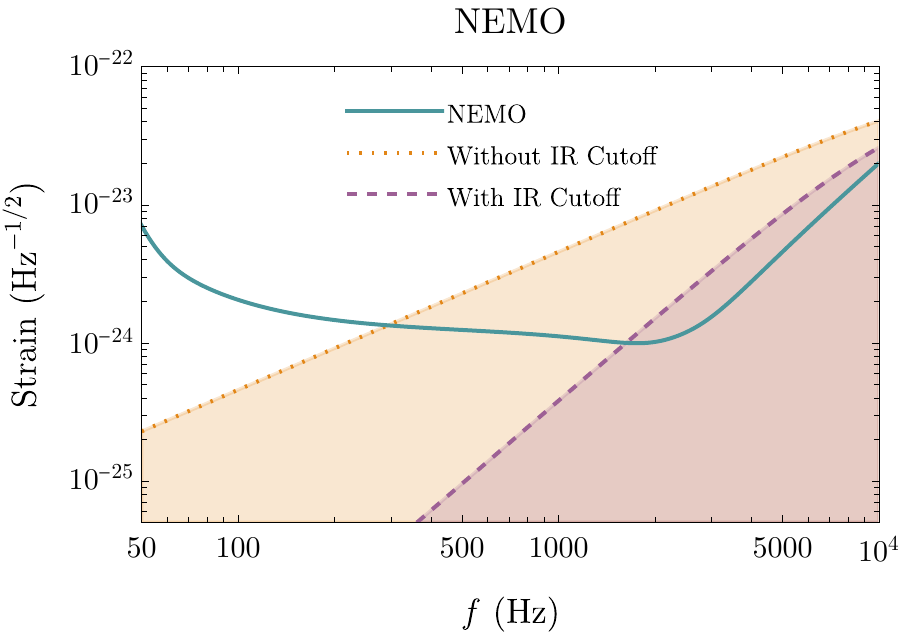}
        }
        \caption{Pixellon strain (dashed and dotted lines) overlaid with the strain sensitivities for LIGO \cite{Lee_LIGO_2021} and NEMO \cite{NEMO_sensitivity} (solid lines). The LIGO data was obtained from the Livingston detector, and the NEMO data omits suspension thermal noise. The dotted lines give the pixellon strain from Eq.~\eqref{eq:hC} computed without an IR cutoff, and the dashed lines give the same quantity including the IR cutoff from Eq.~\eqref{eq:C_IR_cutoff}. We again compute the pixellon strain with $\alpha = 1$.}
        \label{fig:strain_ligo_nemo}
    \end{figure*}
    
    Ref.~\cite{Li:2022mvy} calculated the auto-correlation of length fluctuations in an L-shaped interferometer due to geontropic fluctuations. In this case, we have $\mathbf{x}_{I}=\mathbf{x}_{II}$ and $\mathbf{n}_{I}=\mathbf{n}_{II}$, so we can set the origin of the coordinates to coincide with the beam splitter of the interferometer. Furthermore, we can align the $x$--$y$ plane with the plane of the interferometer and choose the $x$-axis to be along the first arm of the interferometer. Then the whole configuration is determined by the separation angle $\theta$ between two arms. In this case, the first two terms are the same in Eq.~\eqref{eq:tautau} and similarly for the last two terms, so Eq.~\eqref{eq:tautau} reduces to 
    \begin{equation} \label{eq:C_T_Lshaped_psd}
        \tilde{C}_{\mathcal{T}}(\omega,\theta)
		=2\tilde{C}(\omega,0)-2\tilde{C}(\omega,\theta)\,,
    \end{equation}
    which is consistent with Eq.~(45) of Ref.~\cite{Li:2022mvy}. The spectrum $\tilde{C}(\omega,\theta)$ is given by Eq.~\eqref{eq:C_psd} after setting $L_1=L_2=L$, where $L$ is the length of the interferometer, {\em i.e.},
    \begin{equation} \label{eq:C_Lshaped_psd}
    \begin{aligned}
	\tilde{C}(\omega,\theta)
	=& \;\frac{al_p}{8\pi c_s^3L^2}
	\int_{0}^{L}dr_1\int_{0}^{L}dr_2\;
	\sinc\left[\omega\mathcal{D}(r_1,r_2,\theta)/c_s\right] \\
	& \;\cos{\left[\omega(L-r_1)\right]}
	\cos{\left[\omega(L-r_2)\right]}\,.
    \end{aligned} 
    \end{equation}
    where the distance factor $\mathcal{D}$ is now completely determined by $r_1$, $r_2$, and $\theta$,
    \begin{equation}
       \mathcal{D}(r_1,r_2,\theta)
		=\sqrt{r_1^2+r_2^2-2r_1r_2\cos{\theta}}\,.
    \end{equation}

    To compare against the strain sensitivity of real experiments, one needs to first convert Eq.~\eqref{eq:C_Lshaped_psd} to the one-sided noise strain $S_h$ defined by Refs.~\cite{Moore_2014, Chou_2017}
    \begin{equation} \label{eq:noise_strain_def}
       \sqrt{S_h(f)} = \sqrt{2\int_{-\infty}^{\infty}\left\langle\frac{\Delta L(\tau)}{L}\frac{\Delta L(0)}{L}\right\rangle e^{-2\pi if\tau}d\tau}\,,
    \end{equation}
    which has units of Hz$^{-1/2}$. In many of these interferometers, Fabry-P\'{e}rot cavities are used to increase the sensitivity, in which light travels multiple round trips. By converting the strain sensitivity to the phase sensitivity, Ref.~\cite{Li:2022mvy} showed that the geontropic signal does accumulate in Fabry-P\'{e}rot cavities since the output is linear in the phase shift of the light. Thus, it is legitimate to compare our PSD to the strain sensitivity of these experiments. From Eqs.~\eqref{eq:corr_T_def} and \eqref{eq:noise_strain_def}, Ref.~\cite{Li:2022mvy} found that 
    \begin{equation} \label{eq:hC}
        \sqrt{S_{h}(f)} = \sqrt{2\tilde{C}_{\mathcal{T}}\left(\omega,\theta\right)}\,.
    \end{equation}
    Nonetheless, the signal's shape is determined by the geometry of one light-crossing. For example, we expect that the signal peak is at $\omega L\sim 1$, where $L$ is the length of the interferometer instead of the total distance traveled across multiple light-crossings.

    Using Eqs.~\eqref{eq:C_T_Lshaped_psd}--\eqref{eq:C_Lshaped_psd}, Ref.~\cite{Li:2022mvy} computed the PSD of the pixellon model in several L-shaped interferometers (Holometer~\cite{Chou_2017}, GEO-600~\cite{geo_600}, and LIGO~\cite{Lee_LIGO_2021}) and one set of interferometers in LISA~\cite{LISA_2021}, and compared the signal to their strain sensitivities. It was found that GEO-600 and LISA are unlikely to detect geontropic fluctuations due to their relatively low peak sensitivity (at $\omega L\sim1$), while LIGO and Holometer respectively constrain the $\alpha$-parameter to be $\alpha\lesssim 3$ and $\alpha\lesssim 0.7$ (with an IR cutoff), and $\alpha\lesssim 0.1$ and $\alpha\lesssim 0.6$ (without an IR cutoff) at $3\sigma$ significance. Note that the LIGO sensitivity data that we have used here and in Ref.~\cite{Li:2022mvy} is that from Ref.~\cite{Lee_LIGO_2021} with the quantum shot noise removed ({\em i.e.}, the gray curve in Fig.~2 of Ref.~\cite{Lee_LIGO_2021}) by the quantum-correlation technique in Ref.~\cite{Martynov:2017ufx}. Nonetheless, this technique only removes the expectation value of the shot noise but not its variance \cite{Yu:2022upc}, limiting the extent to which we can dig under the shot noise.  More specifically, with a frequency band of $\Gamma$ and an integration time of $T$, we expect the noise suppression factor to be $\sim (\Gamma T)^{1/4}$ in amplitude --- or until the next underlying noise is revealed.  In the particular case of LIGO, that underlying noise includes coating and suspension thermal noise at low frequencies, and laser noise at high frequencies.  Further studying these underlying noise sources in LIGO can in principle put more stringent upper limits on the geontropic noise. 

    Besides the GW detectors above, there are other future L-shaped interferometers to be considered but not included in Ref.~\cite{Li:2022mvy}. The most important ones are the third-generation GW detectors: CE \cite{Evans:2021gyd, Srivastava:2022slt} and ET \cite{Hild:2010id}. CE is a ground-based broadband GW detector using dual-recycled Fabry-P\'{e}rot Michelson interferometers with perpendicular arms. CE will have two sites with several potential designs: a $20$~km interferometer paired with a $40$~km interferometer, or a pair of $20$~km or $40$~km interferometers. As largely a scale-up of Advanced LIGO \cite{Evans:2021gyd}, CE will operate at room temperature with a fused-silica coating of mirrors to reduce thermal noise, and degenerate optical parametric amplifiers injecting squeezed light with low phase uncertainty to reduce quantum noise (shot noise) at high frequency \cite{Tse:2019wcy}. 
    
    ET is an equilateral triangle configuration of three independent nested detectors, each of which contains two dual-recycled Fabry-P\'{e}rot Michelson interferometers with arms of length $10$~km (plotted in Fig.~\ref{fig:ET}) for low- and high-frequency detections, respectively. ET will be built underground to reduce seismic noise. Cryogenic systems are used to reduce thermal noise by cooling the optical systems to $10$--$20$~K at low frequency, while squeezed light (frequency-dependent) is also inserted to reduce quantum noise at high frequency \cite{ET:design}.
    
    As briefly discussed in Sec.~\ref{sec:introduction} and shown in Fig.~\ref{fig:strain_ce_et}, for the benchmark value $\alpha = 1$, the PSD of the geontropic signal overwhelms the strain sensitivity of CE and ET by about two orders of magnitude for $f\sim 1$~kHz. For CE, we have considered both the interferometers of length $20$~km and $40$~km. For ET, we have computed the auto-correlation of a single interferometer within the entire configuration. A study of the cross-correlation of different interferometers is carried out in Sec.~\ref{sec:triangular}.
    
    Besides ET and CE, another next-generation GW detector is NEMO \cite{Ackley:2020atn}, a Michelson interferometer with perpendicular Fabry-P\'{e}rot arms of length $4$~km. Although with less sensitivity than the full third-generation detectors in general, NEMO is important for testing technological developments to be used in the third-generation detectors while making interesting scientific discoveries, such as understanding the compositions of NSs. Due to its interest in binary NS mergers, NEMO specializes in high-frequency events with its optimal sensitivity at $f\sim 1$--$4$~kHz \cite{Ackley:2020atn}. As plotted in Fig.~\ref{fig:strain_ligo_nemo}, within the optimal sensitivity of NEMO, the geontropic signal exceeds the strain sensitivity by about one order of magnitude. Thus, the geontropic signal must be constrained before these next-generation GW detectors can detect other high-frequency events. For future detectors, we have compared the geontropic signal with their design sensitivities, without considering removal of shot noise via the quantum-correlation approach --- even though at high frequencies, where the constraints for geontropic noise are the best, these detectors are limited by shot noise.  It can be anticipated that at these frequencies, these detectors' shot noise dominates over other types of noise by a significant factor.  In this way, these detectors are capable of putting much more stringent bounds on the geontropic $\alpha$ parameter.

    \subsection{Equilateral triangle configurations}
    \label{sec:triangular}

    \begin{figure}[t] 
	\centering
	\includegraphics[width=1\linewidth]{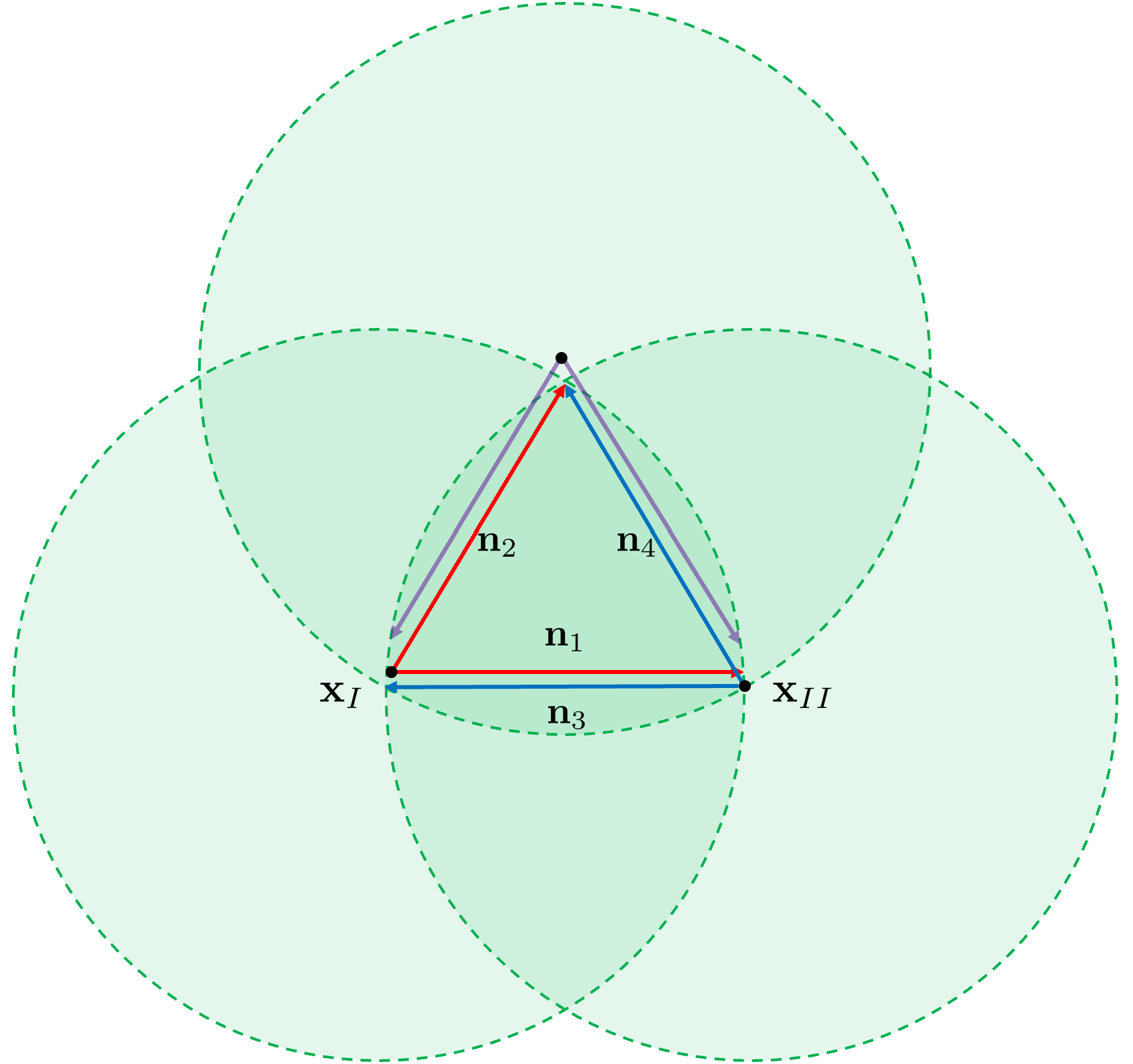}
	\caption{Setup of ET. The red, blue, and purple rays correspond to the three detectors in ET, where we have only shown one of the two interferometers within each detector. We choose not to plot the mirrors at the endpoints of the light beams for simplicity.}
	\label{fig:ET}
    \end{figure}

    In this subsection, we consider configurations of multiple interferometers with certain geometries. For GW detections, these different geometries are helpful in retrieving the polarization of GWs. One important configuration is the equilateral triangle configuration of three interferometer arms, such as LISA \cite{LISA_2021}, or three partially overlapping independent detectors, such as ET \cite{Hild:2010id}, as shown in Fig.~\ref{fig:ET}. For LISA, the signals of different arms can be time shifted and linearly combined to form virtual Michelson interferometers \cite{Freise:2008dk, LISA:2017pwj}. Nonetheless, as found in Ref.~\cite{Li:2022mvy} and discussed in Sec.~\ref{sec:L-shape}, LISA is not promising for detecting geontropic signals, so we will focus on the specific configuration of ET.

    In this subsection, we will study the cross-correlation of multiple detectors of ET. For stochastic wave backgrounds with completely random radiation, a single detector cannot distinguish the background from random instrumental noise within a short observing time unless the sources distribute anisotropically \cite{et_science_team_2011_3911261}. 
    However, since the ET detectors occupy the same spatial region, geontropic fluctuations modeled by the pixellons are correlated between them. Assuming that the noises of different ET detectors are largely uncorrelated, cross-correlating multiple ET detectors allows us to dig under the noise with a suppression factor $\sim (\Gamma T)^{1/4}$, or until a correlated noise background is reached \cite{romano2017detection,et_science_team_2011_3911261}. By contrast, the single-detector quantum-correlation technique discussed in Sec.~\ref{sec:L-shape} only allows us to dig under the shot noise, and it will be limited by non-quantum noise sources of a single detector. This motivates the calculation of cross-correlations of different detectors within configurations of interferometers, such as ET. 
    
\begin{figure}[t] 
		\centering
		\includegraphics[width=1\linewidth]{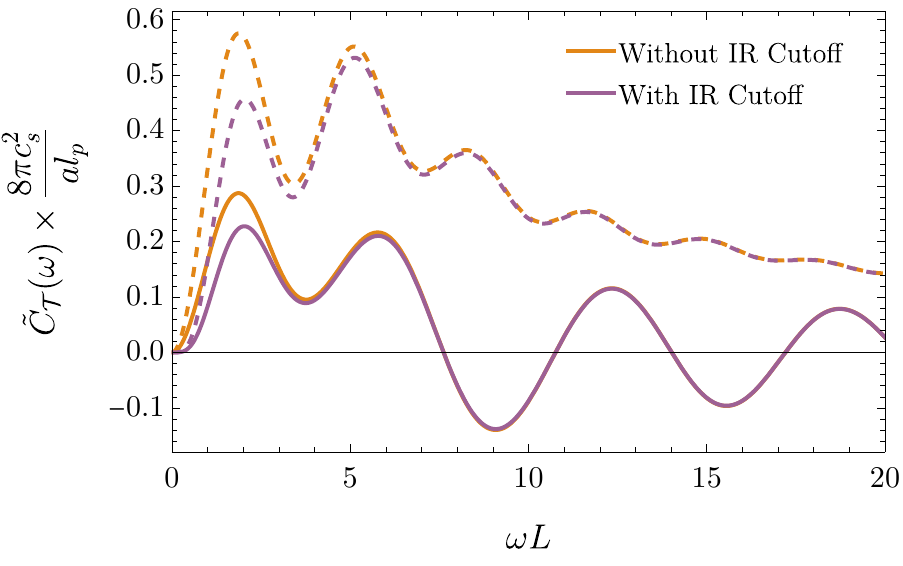}
		\caption{The PSD $\tilde{C}_{\mathcal{T}}(\omega)$ of the cross-correlation function of two sets of interferometers within a triangular configuration like ET [Eq.~\eqref{eq:PSD_cross_ET}, solid lines], together with the corresponding auto-correlation $\tilde{C}_{\mathcal{T}}(\omega, \theta=\frac{\pi}{3})$ of a single interferometer within this configuration [Eq.~\eqref{eq:C_T_Lshaped_psd}, dashed lines].}
		\label{fig:PSD_ET}
    \end{figure}

\begin{figure}[t]
        \centering
        \includegraphics[width=1\linewidth]{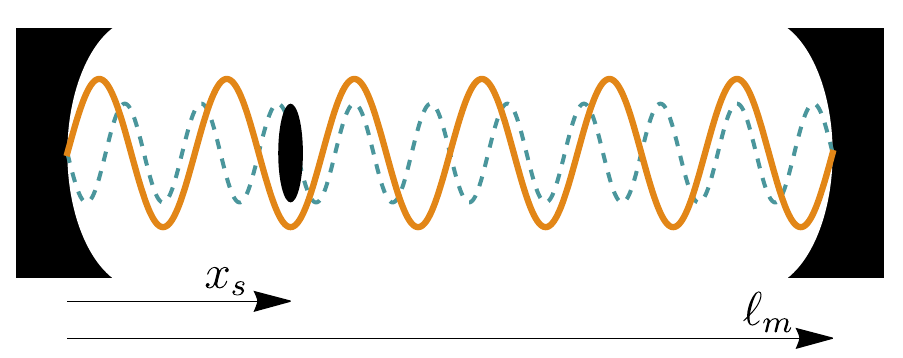}
        \caption{Schematic diagram of the optically-levitated sensor as described in Refs.~\cite{Arvanitaki:2012cn,Aggarwal:2020umq}. A dielectric sphere or microdisk is trapped in an anti-node of an optical cavity (solid orange). A second laser (dashed blue) is used to cool the sensor and read out its position. Transverse motion is cooled by additional lasers (not shown).}
        \label{fig:levitated_sensor}
    \end{figure}

    Let us consider one set of two interferometers across different detectors within ET, {\em e.g.}, the red and blue detectors in Fig.~\ref{fig:ET}, and pick the origin of coordinates at the origin of the red detector $\mathbf{x}_1$. Let us also pick the $x$--$y$ plane to be the plane of the interferometers, with the $x$-axis along $\mathbf{n}_1$. In this case,
    \begin{align}
        & \mathbf{x}_I=0\,,\;\mathbf{x}_{II}=L\hat{\mathbf{x}}\,,\;
        \mathbf{n}_1=\hat{\mathbf{x}}\,,\;
        \mathbf{n}_2=\frac{1}{2}\hat{\mathbf{x}}
        +\frac{\sqrt{3}}{2}\hat{\mathbf{y}}\,, \nonumber\\
        & \mathbf{n}_3=-\hat{\mathbf{x}}\,,\;
        \mathbf{n}_4=-\frac{1}{2}\hat{\mathbf{x}}
        +\frac{\sqrt{3}}{2}\hat{\mathbf{y}}\,.
    \end{align}
    Here, we have assumed that the arms along the same line completely overlap with each other ({\em i.e.}, the arms along $\mathbf{n}_1$ and $\mathbf{n}_3$). In reality, there is a finite separation between these arms, which can be dealt with via the general procedure in Sec.~\ref{sec:pixellon_extension}. Then one can compute $\mathcal{D}(r_{i,j},\Delta\mathbf{x},\mathbf{n}_{i,j})$ for all the combinations in Eq.~\eqref{eq:tautau}, {\em i.e.},
    \begin{align} \label{eq:D_ET}
        & \mathcal{D}_{13}(r_{1},r_{2})=|r_1+r_2-L|\,, \nonumber\\
        & \mathcal{D}_{24}(r_{1},r_{2})
        =\frac{1}{2}\sqrt{(2L-r_1-r_2)^2+3(r_1-r_2)^2}\,, \nonumber\\
        & \mathcal{D}_{14}(r_{1},r_{2})
        =\frac{1}{2}\sqrt{(2L-2r_1-r_2)^2+3r_2^2}\,, \nonumber\\
        & \mathcal{D}_{32}(r_{1},r_{2})
        =\mathcal{D}_{14}(r_{1},r_{2})\,.
    \end{align}
    Here, we have defined $\mathcal{D}_{ij}(r_{1},r_{2})$ such that $r_1$ is the integration variable along the arm with direction $\mathbf{n}_{i}$, and $r_2$ is the integration variable along the arm with direction $\mathbf{n}_{j}$. Plugging Eq.~\eqref{eq:D_ET} into Eq.~\eqref{eq:tautau}, we get
    \begin{equation} \label{eq:PSD_cross_ET}
        \begin{aligned}
            \tilde{C}_{\mathcal{T}}(\omega)=
            & \;\frac{al_p}{8\pi c_s^3L}
		  \int_{0}^{L}dr_1\int_{0}^{L}dr_2\; \\
		  & \;\cos{\left[\omega(L-r_1)\right]}
		  \cos{\left[\omega(L-r_2)\right]} \\
		  & \;\left\{\sinc\left[
            \omega\mathcal{D}_{13}(r_{1},r_{2})/c_s\right]
            +\sinc\left[\omega\mathcal{D}_{24}(r_{1},r_{2})/c_s\right]\right. \\
            & \;\left.-2\sinc\left[\omega\mathcal{D}_{14}(r_{1},r_{2})/c_s\right]\right\}\,,
        \end{aligned}
    \end{equation}
    the result of which is plotted in Fig.~\ref{fig:PSD_ET}.

    Besides the equilateral triangle configuration of ET, one can compute the response of other geometries of interferometers to the pixellon model following the procedure in Sec.~\ref{sec:pixellon_extension}. For example, one can consider two or multiple interferometers with the same length located at the same origin but rotated from each other by certain angles as depicted in Ref.~\cite{Freise:2008dk}. There are even more complicated geometries, such as the twin 3-d interferometers that will be built at Cardiff University \cite{Vermeulen:2020djm}. The authors in Ref.~\cite{Vermeulen:2020djm} claimed that the angular correlations of geontropic fluctuations, as discussed in detail in Refs.~\cite{Verlinde_Zurek_2019_1, Zurek_2020, Verlinde_Zurek_3, Li:2022mvy}, especially the transverse correlations due to the low-$\ell$ modes, can be probed by this geometry. While, in principle, the geontropic signal can be computed for such a complicated interferometer geometry, the pixellon model may not adequately encapsulate the underlying physics of the VZ effect.  Further, first-principles calculations of geontropic fluctuations assume a simple causal diamond radiating outward from a beam splitter.  One major feature of the twin 3-d interferometers in Ref.~\cite{Vermeulen:2020djm} is that the interferometer arms are bent at mirrors $\mathrm{MM_{A}}$ and $\mathrm{MM_{B}}$ (see Fig.~1 of Ref.~\cite{Vermeulen:2020djm}), so the causal diamond of the whole apparatus is distorted. The bent-arm configuration explicitly breaks spherical symmetry, which the previous calculations \cite{Zurek_2020,Li:2022mvy} relied on. Specifically, the pixellon metric in Eq.~\eqref{eq:metric_pix} captures metric fluctuations only along interferometer arms that extend radially outward from a beam splitter. One can decompose the bent interferometer arms into segments of straight arms, and, assuming the pixellon model pertains to such a causal diamond, attempt to apply the pixellon model to each segment by choosing \textit{local} coordinates centered at the beam splitter, $\mathrm{MM_{A}}$, and $\mathrm{MM_{B}}$, respectively. However, the major obstacle for this procedure is that at $\mathrm{MM_{A}}$ (or $\mathrm{MM_{B}}$) there does not exist a closed causal diamond, because light continues to traverse past $\mathrm{MM_{A}}$ (or $\mathrm{MM_{B}}$) until it reaches $\mathrm{EM_{A}}$ (or $\mathrm{EM_{B}}$) or the beam splitter. Since the calculations in Refs.~\cite{Zurek_2020,Li:2022mvy} require a closed causal diamond such that the observable computed is manifestly gauge invariant, one first needs to ascertain whether the procedures in Ref.~\cite{Li:2022mvy} for computing gauge-invariant quantities are still valid when piecing together these non-closed causal diamonds. Due to these complications, we do not attempt to apply the pixellon model to the Cardiff experiment, as we believe that an accurate prediction for such bent-arm configurations will require a more direct, first-principles calculation requiring better theoretical control than current technology allows. In the next subsection, we focus on another interferometer-like experiment, the optically-levitated sensor. 

    \subsection{Optically-levitated sensor}
    \label{sec:levitated_sensor}

    In this subsection, we study the response of the optically-levitated sensor in Refs.~\cite{Arvanitaki:2012cn, Aggarwal:2020umq} to geontropic fluctuations described by the pixellon model. To understand the working principle of the optically-levitated sensor, let us first consider its response to GWs following Ref.~\cite{Arvanitaki:2012cn}, working in the local Lorentz frame with origin at the input mirror. Let the unperturbed distance between the optical cavity mirrors be $\ell_m$, and the unperturbed distance from the input mirror to the sensor in its trap minimum be $x_s$. Under a passing GW perpendicular to the cavity with strain $h$, the proper distances to the mirror and sensor are both shifted,
    \begin{equation}
        \delta x_s=\frac{1}{2}hx_s\,,\quad
        \delta \ell_m=\frac{1}{2}h\ell_m\,.
    \end{equation}
    The new position of the trap minimum can be found from the condition
    \begin{equation}
        k_t(\ell_m'-x_\mathrm{min}')=k_t(\ell_m-x_\mathrm{min})
        =\left(n+\frac{1}{2}\right)\pi\,,
    \end{equation}
    where $n$ is an integer, and $k_t$ is the wavenumber of the trapping laser. The shift of the trap minimum is then given by $\delta x_\mathrm{min}=\ell_m'-\ell_m=\delta \ell_m$. Here, we have assumed that the trapping laser has a constant frequency inside the cavity. Thus, the sensor is displaced from its trap minimum by an amount given in Ref.~\cite{Arvanitaki:2012cn} as
    \begin{equation}
        \Delta X\equiv\delta x_s-\delta x_{\min}
        =\frac{1}{2}h(x_s-\ell_m)+\mathcal{O}(h^2)\,.
        \label{eq:trap_displacement}
    \end{equation}
    This displacement will result in an oscillatory driving force on the sensor. If the GW frequency matches the trapping frequency $\omega_0$ of the sensor, the driving force will resonantly excite the sensor. The corresponding oscillations can then be measured. When $x_s\ll\ell_m$, the effect of the GW is maximized.

    \begin{figure}[t]
        \centering
        \includegraphics[width=1\linewidth]{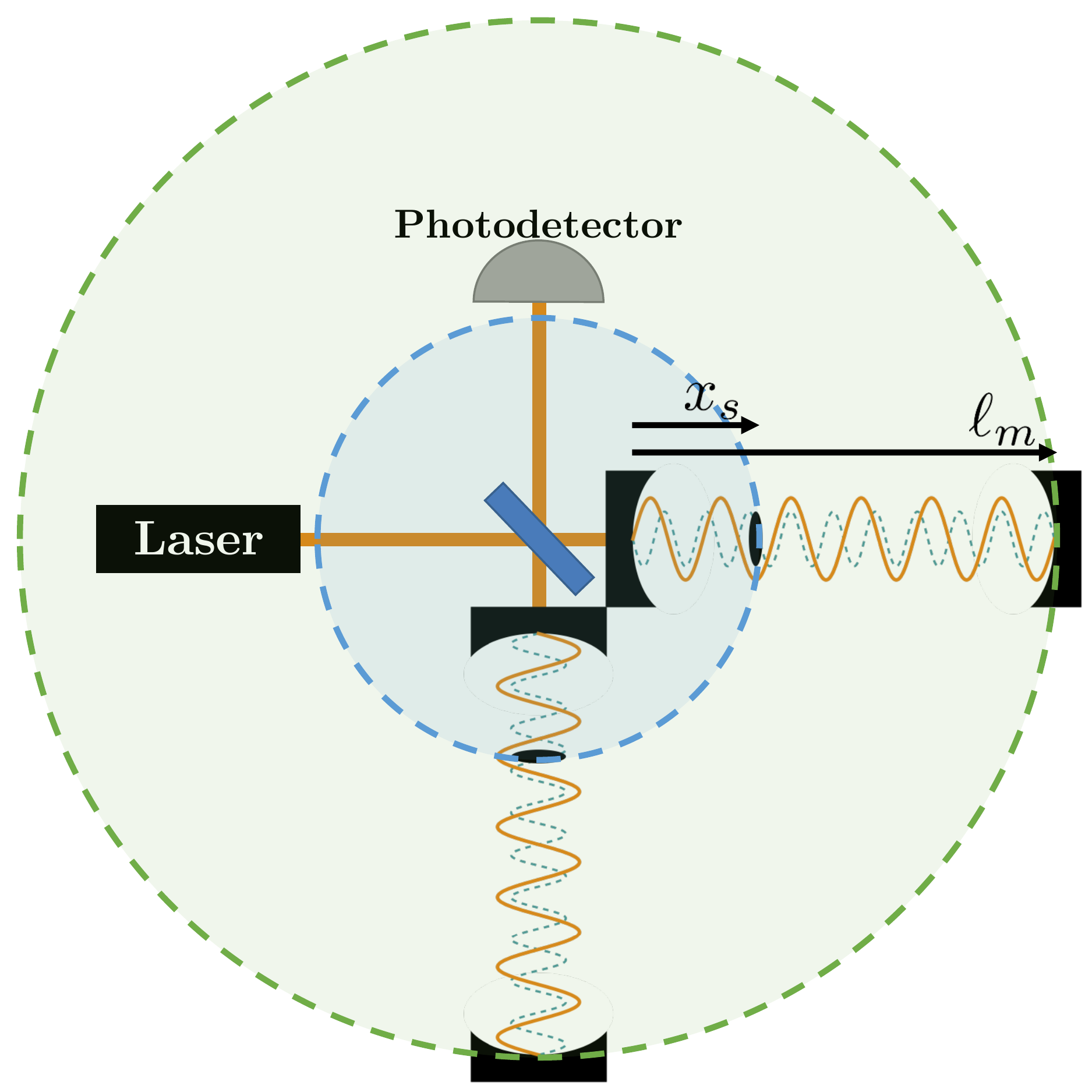}
        \caption{Two levitated sensors inserted into the Fabry-P\'{e}rot cavities of a Michelson interferometer, as described in Ref.~\cite{Aggarwal:2020umq}. The entangling surfaces corresponding to the two arms of length $x_s$ and $\ell_m$ are marked by the blue and green shaded circles, respectively. Note that this diagram ignores the distances between the beam splitter and the input mirrors of the two cavities.}
        \label{fig:levitated_sensor_interferometer}
    \end{figure}

    For the pixellon model, the response of the optically-levitated sensor can be calculated similarly. In our case, $\delta x_s$ and $\delta \ell_m$ are given by
    \begin{align}
        \delta x_s=-\frac{1}{4}\int_0^{x_s}dr\left[\phi(x)+\phi(x')\right]\,, \label{eq:Xs_roundtrip} \\
        \delta \ell_m=-\frac{1}{4}\int_0^{\ell_m}dr\left[\phi(y)+\phi(y')\right]\,, \label{eq:Lm_roundtrip}
    \end{align}
    where
    \begin{equation}
    \begin{alignedat}{2}
        x &= (t_x-x_s+r, r\mathbf{n})\,,\quad
        x' &&= (t_x+x_s-r, r\mathbf{n})\,,  \\
        y &= (t_\ell-\ell_m+r, r\mathbf{n})\,,\quad
        y' &&= (t_\ell+\ell_m-r, r\mathbf{n})\,,
    \end{alignedat}
    \end{equation}
    and the start times of each beam are chosen to be $t_x-x_s$ and $t_\ell-\ell_m$. Note the additional factor of $\frac{1}{2}$ as compared to Eq.~\eqref{eq:time_delay_total}, since the lengths $\ell_m$ and $x_s$ are one-half of the corresponding round-trip time delays when there are no geontropic fluctuations. Within a single arm, since there is only a single beam measuring the position of the sensor, we can choose 
    \begin{equation}
        t_x=t+x_s\,,\quad t_\ell=t+\ell_m
    \end{equation}
    such that the start times of the beam probing the sensor and the end mirror are the same. Notice that, in general, two independent pixellon models should be used for the shorter and longer arms. Nevertheless, since both spherical entangling surfaces are located at the same origin, as depicted in Fig.~\ref{fig:levitated_sensor_interferometer}, and the pixellon fields $\phi$ are universal across these two causal diamonds as discussed in Sec.~\ref{sec:pixellon_extension}, the forms of Eqs.~\eqref{eq:Xs_roundtrip} and \eqref{eq:Lm_roundtrip} are very similar. This is consistent with the fact that the metric in Eq.~\eqref{eq:metric_pix} is spatially conformal.
    
    The displacement of the levitated sensor from its trap minimum is then given by
    \begin{equation}
        \begin{aligned} \label{eq:DeltaX}
        \Delta X
        =& \;-\frac{1}{4}\int_0^{x_s}dr\left[\phi(x)+\phi(x')\right]
        +\frac{1}{4}\int_0^{\ell_m}dr\left[\phi(y)+\phi(y')\right].
        \end{aligned}
    \end{equation}
    Note that Eq.~\eqref{eq:DeltaX} is similar, but not identical to, the round-trip time of a photon traveling from position $x_s$ to $\ell_m$, {\em i.e.},
    \begin{equation} \label{eq:DeltaX_alt}
        \begin{aligned}
        & \Delta X |_{x_s 
        \leftrightarrow \ell_m}
        =\frac{1}{4}\int_{x_s}^{\ell_m}dr
        \left[\phi(y)+\phi(y')\right]\,, \\
        & y=(t-\ell_m+r, r\mathbf{n})\,,\quad
        y'=(t+\ell_m-r, r\mathbf{n}).
        \end{aligned}
    \end{equation}
    Using Eq.~\eqref{eq:DeltaX_alt} instead of Eq.~\eqref{eq:DeltaX} would give a PSD identical to Eq.~\eqref{eq:C_Lshaped_psd} with length $L=\ell_m-x_s$.
    
    We can then define the correlation function of $\Delta X$ as
    \begin{equation}
        C^{\Delta X}(\Delta t,\theta)
        \equiv\left\langle\frac{\Delta X(t_1,\mathbf{n}_1)
        \Delta X(t_2,\mathbf{n}_2)}{(\ell_m -x_s)^2}\right\rangle\,,
    \end{equation}
    where the unit vectors $\mathbf{n}_i$ parameterize the orientations of the two levitated sensor arms, and the angle $\theta$ between them is given by $\cos(\theta)=\mathbf{n}_1\cdot\mathbf{n}_2$. The difference between the beam start times is $\Delta t\equiv t_1-t_2$. Note that the normalization of $C^{\Delta X}$ assumes that the characteristic length of the system is $\ell_m - x_s$, as per the above discussion. Using Eq.~\eqref{eq:DeltaX}, we find that
    \begin{align}
        & C^{\Delta X}(\Delta t,\theta) \\
        =& \;\frac{1}{16(\ell_m-x_s)^2}
        \bigg[\int_0^{x_s} dr_1 \int_0^{x_s} dr_2\;
        \mathcal{C}(x_1,x_2) \nonumber\\
        & \;-\int_0^{x_s} dr_1 \int_0^{\ell_m} dr_2\;
        \mathcal{C}(x_1,y_2)
        -\int_0^{\ell_m} dr_1 \int_0^{x_s} dr_2\; 
        \mathcal{C}(y_1,x_2) \nonumber\\
        & \;+\int_0^{\ell_m} dr_1 \int_0^{\ell_m} dr_2\;
        \mathcal{C}(y_1,y_2)\bigg]\,,
    \end{align}
    where $\mathcal{C}(x,y)$ is defined in Eq.~\eqref{eq:phi_corr2}. The first and last terms above are correlations between the arms with the same length (either $L=x_s$ or $L=\ell_m$). In contrast, the second and third terms correlate arms with different lengths, {\em i.e.}, the arm of $L=x_s$ with the arm of $L=\ell_m$. 
    
    Following a similar calculation as the one to obtain Eq.~\eqref{eq:C_psd}, we find the two-sided PSD $\tilde{C}^{\Delta X}(\omega,\theta)$ as
    \begin{equation} \label{eq:C_levitated_psd}
    \begin{aligned}
        \tilde{C}^{\Delta X}(\omega,\theta)
        =& \;\left[\tilde{C}^{\Delta X}(\omega,x_1,x_2)
        +\tilde{C}^{\Delta X}(\omega,y_1,y_2)\right. \\
        &\; \left.-2\tilde{C}^{\Delta X}(\omega,x_1,y_2)\right]\,,
    \end{aligned}
    \end{equation}
    where the first two terms are given by Eq.~\eqref{eq:C_psd} with $N=(\ell_m-x_s)^2$ and $\mathcal{D}(r_1,r_2,\theta)=\sqrt{r_1^2+r_2^2-2r_1r_2\cos(\theta)}$. The last term, which corresponds to the correlation between the arms of length $L=x_s$ and $L=\ell_m$, carries an additional geometrical factor of $\cos{\left[\omega(\ell_m-x_s)\right]}$ due to the difference in the sizes of the causal diamonds, {\em i.e.},
    \begin{equation} \label{eq:C_xy_psd}
        \begin{aligned}
		  & \tilde{C}^{\Delta X}(\omega,x_1,y_2) \\
		  =& \;\frac{al_p}{8\pi c_s^3(\ell_m-x_s)^2}
		  \int_{0}^{x_s}dr_1\int_{0}^{\ell_m}dr_2\;
		  \cos{\left[\omega(x_s-r_1)\right]}\\
		  & \cos{\left[\omega(\ell_m-r_2)\right]}
            \cos{\left[\omega(\ell_m-x_s)\right]}
            \sinc\left[\omega\mathcal{D}(r_1,r_2,\theta)/c_s\right]\,.
	\end{aligned}
    \end{equation}
    We can also define $\tilde{C}^{\Delta X}_\mathcal{T}(\omega, \theta)$ as in Eq.~\eqref{eq:C_T_Lshaped_psd} via
    \begin{equation} \label{eq:C_T_levitated_psd}
        \tilde{C}^{\Delta X}_\mathcal{T}(\omega,\theta)
        =2\left[\tilde{C}^{\Delta X}(\omega, 0)
        -\tilde{C}^{\Delta X}(\omega,\theta)\right]\,.
    \end{equation}
    
\begin{figure*}[t]
        \centering
        \subfloat[Pixellon PSD with $x_s = \ell_m / 50$.]{
            \label{fig:psd_comparison_lm50}
            \includegraphics[width=0.48\linewidth]{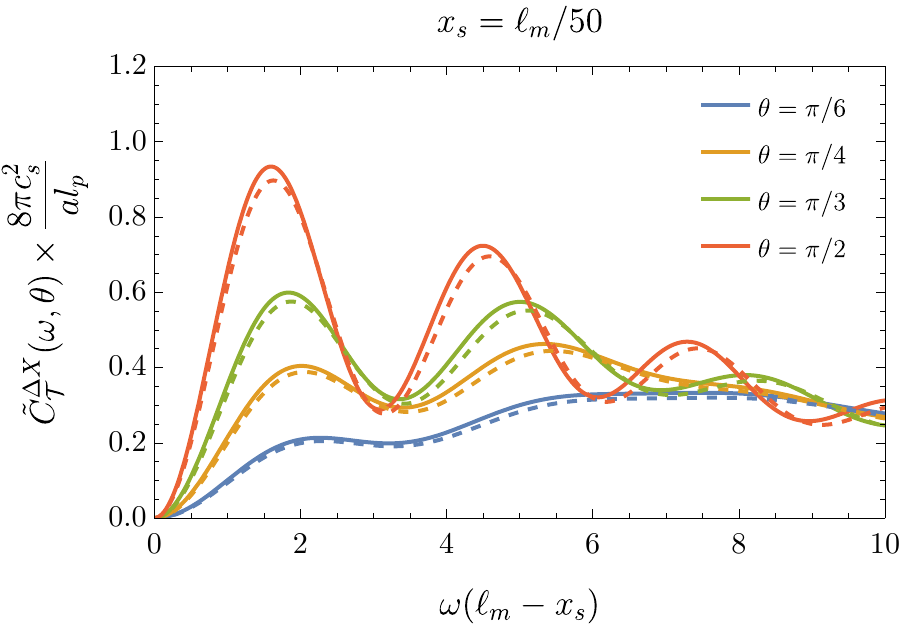}
        }\hfill
        \subfloat[Pixellon PSD with $x_s = \ell_m / 10$.]{
            \label{fig:psd_comparison_lm10}
            \includegraphics[width=0.48\linewidth]{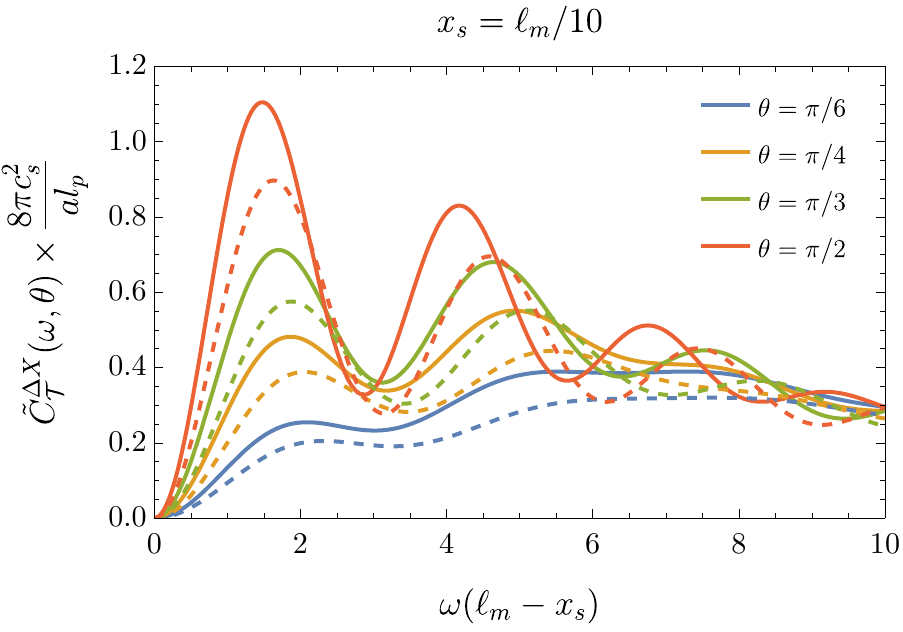}
        }
        \caption{Pixellon PSD $\tilde{C}^{\Delta X}_\mathcal{T}(\omega,\theta)$ as it would appear in an optically-levitated sensor [Eq.~\eqref{eq:C_T_levitated_psd}, solid lines] shown alongside the PSD of an ordinary L-shaped interferometer $\tilde{C}_\mathcal{T}(\omega,\theta)$ [Eq.~\eqref{eq:C_T_Lshaped_psd}, dashed lines]. We take the length of the L-shaped interferometer to be $L = \ell_m - x_s$. All PSDs are computed without an IR cutoff.}
        \label{fig:levitated_sensor_psd_comparison}
    \end{figure*}

    In the limit $x_s\to 0$, only the second term in Eq.~\eqref{eq:C_levitated_psd} is nonzero, corresponding to the length fluctuations of an interferometer of size $L=\ell_m$. Thus, the levitated sensor can be treated as an ordinary interferometer when $x_s$ is sufficiently small. This is confirmed by Fig.~\ref{fig:psd_comparison_lm50}, where we plot the interferometer PSD from Eq.~\eqref{eq:C_Lshaped_psd} against the levitated sensor PSD from Eq.~\eqref{eq:C_levitated_psd}, setting $x_s=\ell_m/50$ and neglecting the IR cutoff for the purpose of demonstration. The interferometer PSD is given by the dashed lines, whereas the levitated sensor PSD is given by the solid lines. We can see that, as expected, the PSDs of these two different types of experiments are very similar in the limit of small $x_s$. In Fig.~\ref{fig:psd_comparison_lm10}, we show a similar comparison but instead pessimistically set $x_s=\ell_m/ 10$. For this larger value of $x_s$, the PSD for the levitated sensor becomes somewhat larger in magnitude compared to that of the ordinary interferometer, but retains a similar shape. In the limit of $\omega\xrightarrow{}0$, we have 
    \begin{equation}
        \tilde{C}^{\Delta X}_\mathcal{T}(\omega,\theta)=\frac{al_p}{48\pi {c_s^5}}\omega^2{{(l_m+x_s)}^2}(1-\cos{\theta})+\mathcal{O}(\omega^4).
    \end{equation}
    From the scaling $\tilde{C}^{\Delta X}_\mathcal{T}(\omega,\theta)\propto {(l_m+x_s)}^2$, one can see the increase of signal as $x_s$ increases, which is a result of treating the system as two sets of causal diamonds. However, we expect the above treatment to break down beyond the limit of $x_s\ll l_m$. We emphasize that this calculation is not intended to be fully rigorous, but rather seeks to provide a heuristic description of the pixellon model in a levitated sensor experiment. Nevertheless, we continue to expect that the levitated sensor will behave similarly to an L-shaped interferometer in the limit of small $x_s$. 
    
    Next, let us compare the PSD found above to the predicted strain sensitivity of optically-levitated sensor experiments. The thermal-noise-limited minimum detectable strain of the optically-levitated sensor at temperature $T_\mathrm{CM}$ is given by Refs.~\cite{Arvanitaki:2012cn, Aggarwal:2020umq} as
    \begin{equation} \label{eq:levitated_sensor_h_limit}
        h_\mathrm{limit}
        =\frac{4}{\omega_0^2 \ell_m}\sqrt{\frac{k_B T_\mathrm{CM}\gamma_g b}{M} \left[1+\frac{\gamma_\mathrm{sc}+R_+}{N_i\gamma_g}\right]}H(\omega_0)\,,
    \end{equation}
    where $\omega_0$ is the trapping frequency, $\gamma_g$ is the gas-damping coefficient, $\gamma_\mathrm{sc}$ is the scattered photon-recoil heating rate, $b$ is the bandwidth, $M$ is the mass of the sensor, and $N_i = k_B T_\mathrm{CM}/\hbar \omega_0$ is the mean initial phonon occupation number. The cavity response function is $H(\omega)=\sqrt{1 +(2\mathcal{F}/\pi)^2\sin^2(\omega \ell_m/c)}$, where $\mathcal{F}$ is the finesse of the cavity. Detailed expressions for all of these quantities can be found in Refs.~\cite{Arvanitaki:2012cn, Aggarwal:2020umq}.

    The peak frequency response of the experiment occurs at the trapping frequency $\omega_0$, at which oscillations of the levitated sensor are resonantly enhanced. The trapping frequency can be widely tuned via the laser intensity \cite{Aggarwal:2020umq}. Thus, the sensitivity curve for the levitated sensor can be obtained by continuously varying the locus of the sensitivity curve for each fixed value of $\omega_0$, as given by Eq.~\eqref{eq:levitated_sensor_h_limit}.

\begin{figure*}[t]
        \centering
        \subfloat[Strain without an IR cutoff.]{
            \label{fig:strain_levitated_sensor_no_IR}
            \includegraphics[width=0.48\linewidth]{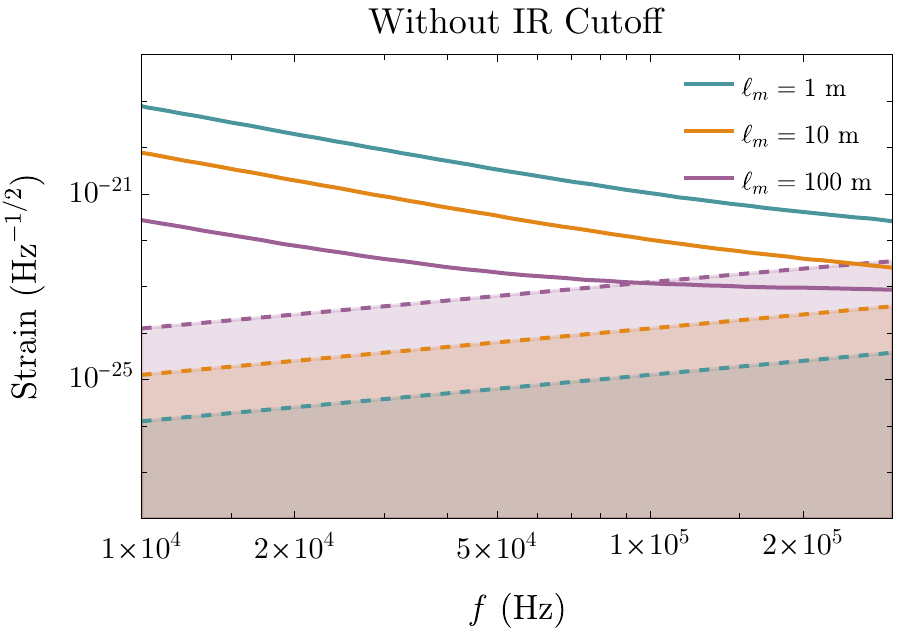}
        }\hfill
        \subfloat[Strain with an IR cutoff $\omega_\mathrm{IR} = 1/(\ell_m - x_s)$.]{
            \label{fig:strain_levitated_sensor_IR}
            \includegraphics[width=0.48\linewidth]{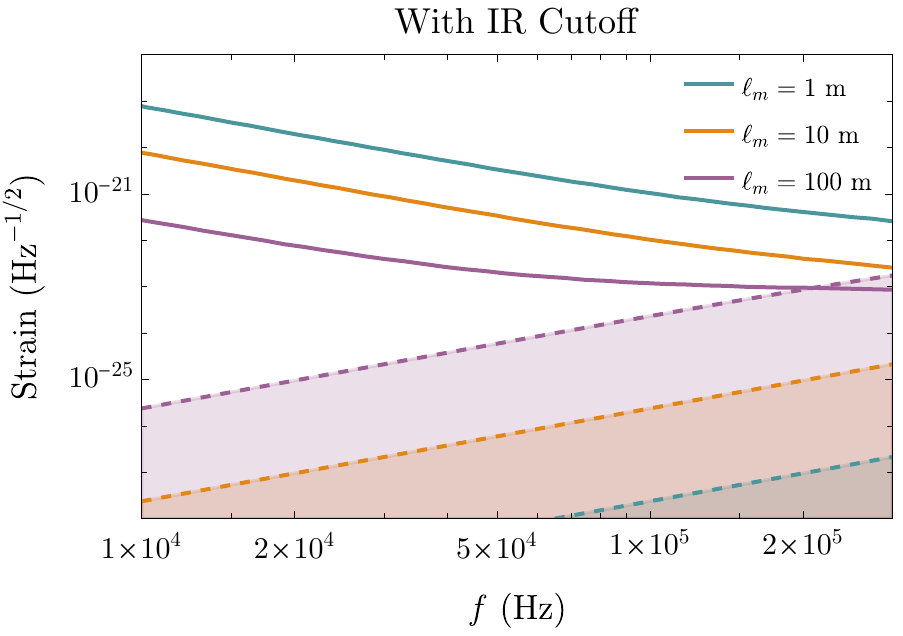}
        }
        \caption{The pixellon strain (dashed lines) overlaid with the predicted strain sensitivity for a stacked-disk levitated sensor (solid lines), as given by Fig.~3 of Ref.~\cite{Aggarwal:2020umq}. The color coding corresponds to the size $\ell_m$ of the levitated sensor. The pixellon strain is computed from Eq.~\eqref{eq:C_T_levitated_psd}, and we set $x_s = \ell_m / 10$ throughout.}
        \label{fig:strain_levitated_sensor}
    \end{figure*}

    In Fig.~\ref{fig:strain_levitated_sensor}, we plot the strain sensitivity of the levitated sensor experiment from Ref.~\cite{Aggarwal:2020umq} (with a sensor consisting of a stack of dielectric disks) against the PSD of the pixellon model from Eqs.~\eqref{eq:C_levitated_psd}--\eqref{eq:C_T_levitated_psd}. In Fig.~\ref{fig:strain_levitated_sensor_IR}, we additionally include an IR cutoff $\omega_\mathrm{IR}=1/L$ as in Eq.~\eqref{eq:C_IR_cutoff}, where we take the characteristic length of the system to be $L = \ell_m - x_s$. This choice comes from the comparison of the displacement $\Delta X$ with the length fluctuations of an interferometer of size $\ell_m - x_s$, as discussed with relation to Eq.~\eqref{eq:DeltaX_alt}. Note that Ref.~\cite{Aggarwal:2020umq} uses a 300 kHz upper bound for their sensitivity curves, citing limitations of power absorption by the suspended sensor. From these plots, we observe that the levitated sensor would only be competitive for detecting the geontropic signal at $\ell_m \gtrsim 100~\mathrm{m}$. At the time of writing, a 1~m prototype of this experiment is under construction, and a 100~m device is at the concept stage \cite{Aggarwal:2020olq,Aggarwal:2020umq}. That these proposed levitated sensor experiments are not competitive for constraining the pixellon model is expected: their reach in frequency is such that $\omega \ell_m \ll 1$, whereas the pixellon signal is expected to peak at $\omega \ell_m \sim 1$.  Finally, let us note that, although the levitated sensors do not move along geodesics, but instead have amplified non-geodesic movements, the same amplification factors are applied to motion induced by the noisy thermal force.  In this way, because the device is limited by thermal noise~\cite{Aggarwal:2020olq,Aggarwal:2020umq}, comparing the displacement~\eqref{eq:DeltaX} and the thermal strain~\eqref{eq:levitated_sensor_h_limit}, as if there were no trapping, still leads to the correct thermal-noise-limited sensitivity.

    \section{Conclusions}
    \label{sec:conclusions}
    
    We have considered the effect of the geontropic signal, from the VZ effect proposed in Refs.~\cite{Verlinde_Zurek_2019_1, Verlinde_Zurek_2019_2, Banks_2021, Gukov:2022oed, Verlinde_Zurek_3}, specifically as modeled in Refs.~\cite{Zurek_2020, Li:2022mvy}, on next-generation terrestrial GW detectors.  We have found that if GQuEST observes spacetime fluctuations from the pixellon, Cosmic Explorer and the Einstein Telescope will have a large background to astrophysical sources from vacuum fluctuations in quantum gravity with which to contend. On the other hand, LISA and other lower-frequency devices are insensitive to this signal. Note that in making these predictions we have assumed the physical equivalence of the pixellon model with the VZ effect for interferometer observables, the proof of which is still the subject of ongoing first-principles calculations.  Even so, given how large the geontropic signal is expected to be in future GW observatories, our results may inform optimal designs for GW observatories, whether searching for quantum or classical sources of GWs.

    \section{Acknowledgements}
    We thank James Gardner for providing us with the strain sensitivity data of NEMO, Evan Hall for providing us with the strain sensitivity data of Cosmic Explorer, and Vincent S.\ H.\ Lee for sharing his code for making the strain sensitivity plots in Ref.~\cite{Li:2022mvy}. We are supported by the Heising-Simons Foundation ``Observational Signatures of Quantum Gravity'' collaboration grant 2021-2817. The work of KZ is also supported by a Simons Investigator award and the U.S. Department of Energy, Office of Science, Office of High Energy Physics, under Award No. DE-SC0011632. The work of YC and DL is also supported by the Simons Foundation (Award Number 568762), the Brinson Foundation and the National Science Foundation (via grants PHY-2011961 and PHY-2011968).

    \bibliographystyle{apsrev4-1}
    \bibliography{reference}
	
\end{document}